\documentclass[reprint,superscriptaddress,amsmath,amsfonts,amssymb,aps,prb,showkeys]{revtex4-2}

\usepackage[T1]{fontenc}

\usepackage{hyperref}
\usepackage{graphicx}
\usepackage{mathrsfs}

\hypersetup{
  breaklinks = {true},
  citecolor = {blue},
  colorlinks = {true},
  linkcolor = {red},
}

\usepackage{physics}
\usepackage{mathtools}
\usepackage[dvipsnames]{xcolor}
\usepackage{braket} 
\usepackage{upgreek}

\usepackage{ulem}

\newcommand{\x}[1]{\mathrm{#1}}
\usepackage{soul}

\begin{document}

\title{Real-Space Approach to Light-Induced Hall Transport in Disordered Materials}

\
\author{Jorge Martínez Romeral}
\affiliation{Catalan Institute of Nanoscience and Nanotechnology (ICN2), CSIC and BIST, Campus UAB, Bellaterra, 08193 Barcelona, Spain}
\affiliation{Department of Physics, Campus UAB, Bellaterra, 08193 Barcelona, Spain}
\author{Luis M. Canonico}
\email{luis.canonico@icn2.cat}
\affiliation{Catalan Institute of Nanoscience and Nanotechnology (ICN2), CSIC and BIST, Campus UAB, Bellaterra, 08193 Barcelona, Spain}
\author{Aron W. Cummings}

\affiliation{Catalan Institute of Nanoscience and Nanotechnology (ICN2), CSIC and BIST, Campus UAB, Bellaterra, 08193 Barcelona, Spain}

\author{Stephan Roche}
\affiliation{Catalan Institute of Nanoscience and Nanotechnology (ICN2), CSIC and BIST, Campus UAB, Bellaterra, 08193 Barcelona, Spain}
\affiliation{ICREA--Instituci\'o Catalana de Recerca i Estudis Avan\c{c}ats, 08010 Barcelona, Spain}

\date{\today}

\begin{abstract}

We introduce a linear-scaling real-space methodology to compute time-resolved electrical responses of materials driven far from equilibrium, with energy relaxation and disorder treated on equal footing. Applying this approach to gapped monolayer and AB-stacked (Bernal) bilayer graphene, when driven by a circularly polarized optical pulse, we observe the generation/suppression of a finite Hall conductivity when the system is trivial/topological. This Hall signal oscillates during optical driving and remains sizable after the light is switched off before relaxing toward equilibrium. Remarkably, this dynamical Hall response is robust in the presence of realistic descriptions of disorder, suggesting that disorder and relaxation dynamics can be leveraged as design parameters rather than as limitations. More broadly, our new methodology enables the investigation of electrical responses in driven, complex disordered quantum materials and highlights how engineered energy-transfer pathways can enable ultrafast optoelectronic functionality.

\end{abstract}

\maketitle
\section{Introduction} 

In recent years, the light-matter interaction has become a powerful driver of novel quantum behavior. Intense and ultrafast optical fields can transiently reshape the electronic structure of solids, enabling phases and transport responses that have no equilibrium counterpart \cite{Floquet_engineering_of_strongly_driven_excitons_in_monolayer_tungsten_disulfide, Possible_light_induced_superconductivity_in_K3C60_at_high_temperature, Strongly_correlated_electron--photon_systems, Transition_from_Optically_Excited_to_Intrinsic_Spin_Polarization_in_WSe_2, Light-field-driven_currents_in_graphene_higuchi_heide_ullmann_weber_hommelhoff_2017, Light_induced_emergent_phenomena_in_2D_materials_and_topological_materials_bao_tang_sun_zhou_2021, Band_structure_engineering_and_non_equilibrium_dynamics_in_Floquet_topological_insulators_rudner_lindner_2020, The_2025_Roadmap_to_ultrafast_dynamics:_Frontiers_of_theoretical_and_computational_modelling5}. 
In this context, two-dimensional materials provide a versatile platform for exploring nonequilibrium transport phenomena. Their reduced dimensionality, tunability, and strong light-matter coupling make them ideal candidates for realizing ultrafast, optically driven transport phenomena, including light-induced Hall responses in systems that are topologically trivial at equilibrium \cite{Photovoltaic_Hall_effect_in_graphene_oka_aoki_2009,Light_induced_anomalous_Hall_effect_in_graphene_mciver_schulte_stein_matsuyama_jotzu_meier_cavalleri_2019,Transport_properties_of_nonequilibrium_systems_under_the_application_of_light:_Photoinduced_quantum_Hall_insulators_without_Landau_levels,Irradiated_graphene_as_a_tunable_Floquet_topological_insulator_usaj_perez-piskunow_foa_torres_balseiro_2014,Floquet_topological_insulator_in_semiconductor_quantum_wells_lindner_refael_galitski_2011,Optical_control_over_topological_Chern_number_in_moire_materials}. 

From a theoretical perspective, out-of-equilibrium transport in driven quantum systems has been addressed using a variety of techniques, including Boltzmann-Floquet approaches \cite{Floquet-Boltzmann_equation_for_periodically_driven_Fermi_systems, Controlled_Population_of_Floquet_Bloch_States_via_Coupling_to_Bose_and_Fermi_Baths, Quantized_transport_and_steady_states_of_Floquet_topological_insulators, Steady_states_of_interacting_Floquet_insulators, Signatures_of_Floquet_electronic_steady_states_in_graphene_under_continuous‑wave_mid‑infrared_irradiation_liu_yang_gaertner_huckabee_suslov_refael_nathan_lewandowski_luis_iliya_esin_et_al_2025}, the non-equilibrium Green’s function formalism \cite{Real-Time_$GW$:_Toward_an_Ab_Initio_Description_of_the_Ultrafast_Carrier_and_Exciton_Dynamics_in_Two_Dimensional_Materials, Nonequilibrium_Green’s_Functions_Approach_to_Inhomogeneous_Systems, Ultrafast_dynamics_of_strongly_correlated_fermions_nonequilibrium_Green_functions_and_selfenergy_approximations, Yambo:_an_ab_initio_tool_for_excited_state_calculations,Quench_dynamics_and_Hall_response_of_interacting_Chern_insulators}, time-dependent density functional theory \cite{Octopus_a_computational_framework_for_exploring_light-driven_phenomena_and_quantum_dynamics_in_extended_and_finite_systems_tancogne-dejean_oliveira_andrade_appel_borca_le_breton_buchholz_castro_corni_correa_et_al_2020,Observation_of_Floquet_Bloch_states_in_monolayer_graphene_choi_mogi_de_giovannini_azoury_lv_su_hübener_rubio_gedik_2025}, and explicit real-time propagation of tight-binding Hamiltonians \cite{Real-time_approach_to_the_optical_properties_of_solids_and_nanostructures:_Time_dependent_Bethe_Salpeter_equation,Expeditious_computation_of_nonlinear_optical_properties_of_arbitrary_order_with_native_electronic_interactions_in_the_time_domain,Giant_self-driven_exciton-Floquet_signatures_in_time-resolved_photoemission_spectroscopy_of_MoS<sub>2</sub>_from_time-dependent_GW_approach,Real-time_dynamics_of_the_photoinduced_topological_state_in_the_organic_conductor}. Related real-time approaches have been used to study the emergence of quasi-steady transport in finite, strongly biased or disordered systems \cite{Unambiguous_simulation_of_diffusive_charge_transport_in_disordered_nanoribbons,From_Bloch_Oscillations_to_a_Steady_State_Current_in_Strongly_Biased_Mesoscopic_Devices,Landauer_transport_as_a_quasisteady_state_on_finite_chains_under_unitary_quantum_dynamics,Markov_Inequality_as_a_Tool_for_Linear_Scaling_Estimation_of_Local_Observables,Photoinduced_anomalous_Hall_effect_in_the_interacting_Haldane_model:_Targeting_topological_states_with_pump_pulses}.  These methods have provided deep insights into energy relaxation and light-induced band engineering, and can incorporate disorder effects at different levels of approximation.

Another class of techniques, often referred to as linear-scaling real-space methodologies \cite{Linear_scaling_quantum_transport_methodologies_fan_garcia_cummings_barrios-vargas_panhans_harju_ortmann_roche_2020, KITE}, has proven crucial for describing electronic transport in complex materials and devices. These approaches are computationally efficient, enabling non-perturbative treatments of disorder and access to the experimentally relevant length scales typically required for realistic modeling \cite{Disorder_Induced_Delocalization_in_Magic_Angle_Twisted_Bilayer_Graphene_guerrero_nguyen_romeral_cummings_garcia_charlier_roche_2025, Scaling_of_the_integrated_quantum_metric_in_disordered_topological_phases_romeral_cummings_roche_2025, Exploring_dielectric_properties_in_atomistic_models_of_amorphous_boron_nitride_galvani_hamze_caputo_kaya_dubois_colombo_nguyen_shin_shin_charlier_et_al_2024}. These real-space methods have traditionally been restricted to equilibrium or near-equilibrium dynamics, but recent developments have overcome this limitation by extending them to the non-equilibrium regime \cite{Real_Time_Out_of_Equilibrium_Quantum_Dynamics_in_Disordered_Materials_canonico_roche_cummings_2024}, with carrier dynamics computed in the presence of arbitrary time-dependent external excitations and carrier relaxation. Following this development, a natural question to ask is: \textit{How are the electronic transport properties of a disordered system reshaped when it is driven out of equilibrium?} To answer this question, we further advance this non-equilibrium real-space quantum dynamics framework to study the electrical response of excited carriers.  The main strengths of this approach are threefold: \textit{(i)} it scales linearly with system size, enabling the simulation of systems with millions of orbitals; \textit{(ii)} it naturally incorporates multiple relaxation mechanisms within a unified density-matrix propagation scheme; and \textit{(iii)} it treats disorder fully non-perturbatively, allowing us to explore the intertwined dynamics of disorder, driving, and relaxation on experimental length scales, while providing connection with measurable material properties such as the electrical conductivity.

Using this framework, we investigate light-induced Hall transport in gapped two-dimensional systems driven far from equilibrium, which was first  theoretically proposed and experimentally measured in monolayer graphene \cite{Photovoltaic_Hall_effect_in_graphene_oka_aoki_2009, Light_induced_anomalous_Hall_effect_in_graphene_mciver_schulte_stein_matsuyama_jotzu_meier_cavalleri_2019, Microscopic_theory_for_the_light_induced_anomalous_Hall_effect_in_graphene, Observation_of_Floquet_states_in_graphene_merboldt_et_al_2025, Observation_of_Floquet_states_in_graphene_merboldt_schüler_schmitt_bange_bennecke_karun_gadge_pierz_schumacher_momeni_steil_et_al_2025}.  As a demonstration of our methodology, we first apply our methodology to the topological phase of the Haldane model \cite{HaldaneModel} and to gapped single-layer graphene. We show that the quantized Hall conductivity of a topological system can be tuned using light, while in the case of gapped graphene, which is topologically trivial at equilibrium, we show that light can generate an anomalous Hall conductivity. 

Furthermore, as a concrete and experimentally relevant platform, we focus on AB-stacked (Bernal) bilayer graphene, which offers a couple distinct advantages: its band gap can be tuned in situ via a perpendicular electric field \cite{Direct_observation_of_a_widely_tunable_bandgap_in_bilayer_graphene_zhang_tang_girit_hao_martin_zettl_crommie_shen_wang_2009}, and it is easily obtained via mechanical exfoliation \cite{Infrared_spectroscopy_of_electronic_bands_in_bilayer_graphene} or CVD growth \cite{Controlled_Growth_of_Bilayer_Graphene_on_Space_Confined_Cu_Substrates,Towards_growth_of_pure_AB_stacked_bilayer_graphene_single_crystals,Layer_by_layer_growth_of_bilayer_graphene_single_crystals_enabled_by_proximity_catalytic_activity,Uniform_growth_of_AB_stacked_bilayer_graphene_single_crystal_by_cyclic_segregation/dissolution}. We consider circularly polarized optical pulses with durations of a few tens of femtoseconds and explicitly account for carrier thermalization due to electron-electron scattering. We observe an emergent Hall conductivity, in concordance with previous experiments \cite{Tunable_and_giant_valley_selective_Hall_effect_in_gapped_bilayer_graphene_yin_tan_barcons-ruiz_torre_watanabe_taniguchi_song_hone_koppens_2022}. The Hall response exhibits nontrivial temporal behavior, including oscillations at twice the driving frequency, reflecting the oscillating nature of the effective magnetic field imprinted in the optical pulse \cite{Heterodyne_Hall_effect_in_a_two_dimensional_electron_gas, Time-resolved_Hall_conductivity_of_pulse_driven_topological_quantum_systems}. Additionally, we show that this dynamical Hall response is robust, or can even be enhanced, in the presence of realistic disorder such as substrate-induced electron-hole puddles. Moreover, our methodology allows us to track the system’s evolution after the optical pulse, showing that the Hall response remains sizable until the system relaxes back to equilibrium. This opens a route to designing ultrafast photodetectors with recovery times on the order of hundreds of femtoseconds. Finally, while we focus on the Hall conductivity and bilayer graphene, the methodology presented here can be applied to arbitrarily complex systems with arbitrary driving or quenching protocols, and is fully generalizable to other electrical responses such as charge, spin or orbital conductivities as well as to charge-to-spin or charge-to-orbital conversion.




\begin{figure}[tbh]
\includegraphics[width=1\columnwidth]{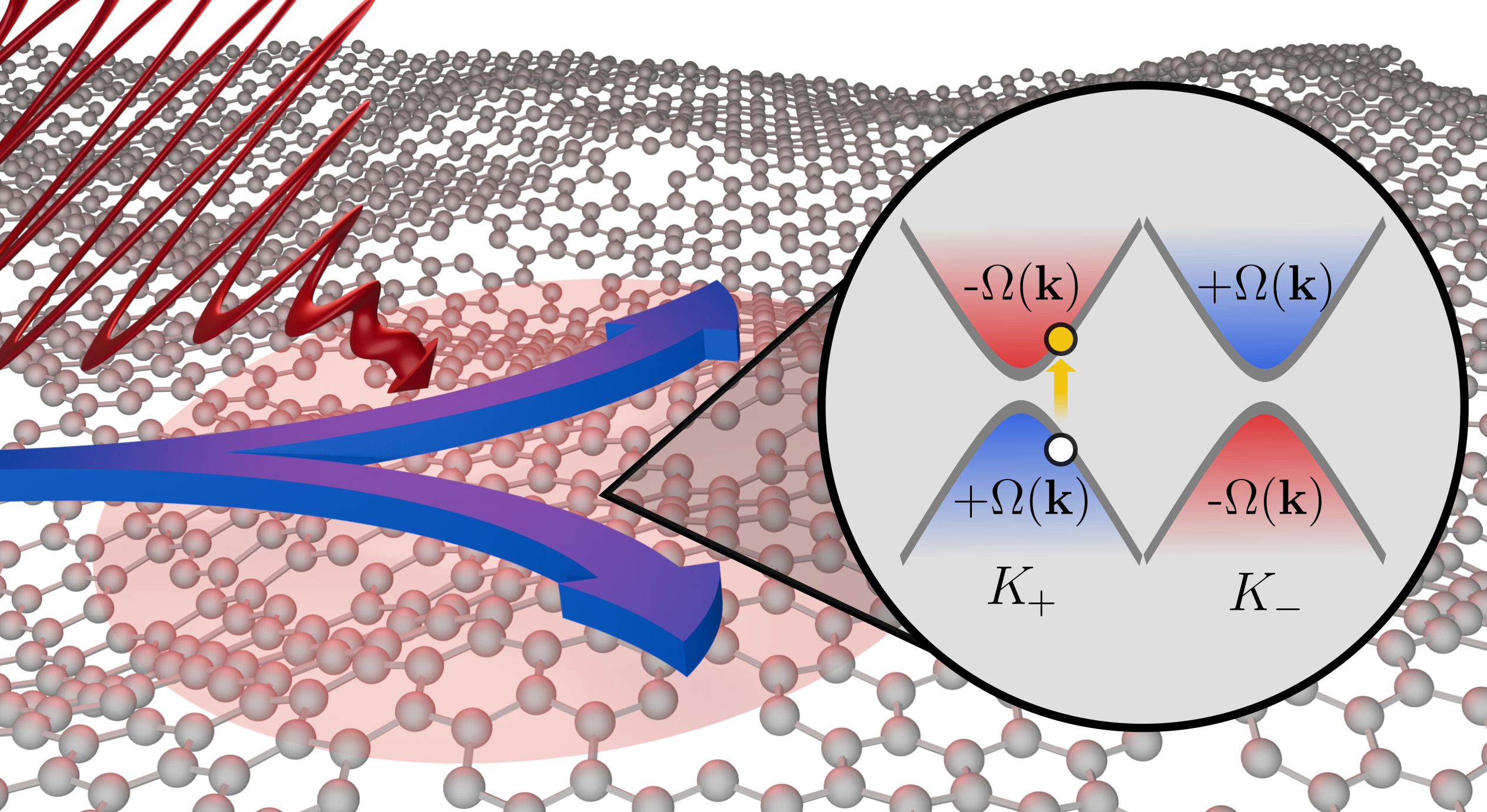}
\caption{Schematic of light-induced quantum transport in disordered graphene, via the valley-selective excitation of carriers from the positive-Berry-curvature valence band to the negative-Berry-curvature conduction band.}
\label{fig:optical_scheme}
\end{figure}

\section{Methodology}
\subsection{Non-equilibrium quantum transport}

Our methodology consists of two steps. First, we simulate the evolution of the density matrix of a given system under a time-dependent Hamiltonian, and second, we calculate observables such as the electrical conductivity and the number of excited carriers at given points in time. 

For the evolution of the density matrix, we follow the approach presented in Ref.\ \cite{Real_Time_Out_of_Equilibrium_Quantum_Dynamics_in_Disordered_Materials_canonico_roche_cummings_2024}, which consists of evolving the ground-state density matrix, $\hat{\rho}(t)$, under the action of a time-dependent Hamiltonian, $\hat{H}(t)$. The system is assumed to initially be in a thermal distribution, $\hat{\rho}(t=0) = \hat{\rho}_{\x{thermal}}(t=0) \equiv \left[1+\exp{\left(\hat{H}(t=0)-\mu(t=0)\right)/k_\x{B} T(t=0)}\right]^{-1}$, with initial temperature $T(t=0)$ and chemical potential $\mu(t=0)$. The system evolves under the time-ordered evolution operator, $\hat{U}(t_1,t_0) = \hat{\cal T} \exp\left\{ -\frac{\x{i}}{\hbar} \int_{t_0}^{t_1} \hat{H}(t') dt' \right\}$, which for short times can be expressed as $\hat{U}(t+\Delta t , t) \approx \exp{(-\x{i}\hat{H}(t)\Delta t/\hbar)}$ \cite{suppmaterial}. Defining the density matrix vector at the initial time, $\ket{\rho(t=0)} \equiv \hat{\rho}(t=0)\ket{\psi}$, its evolution is given by
\begin{align}\label{eq:neq_evolution}
    \ket{\rho(t+\Delta t)} &= \left(1 - \sum_i\dfrac{\Delta t}{\tau_i}\right)\hat{U}(t+\Delta t , t)\ket{\rho(t)}\notag \\
    &+\sum_i \dfrac{\Delta t}{\tau_i}\hat{\rho}_i^{\text{eq}}(t + \Delta t)\hat{U}(t+\Delta t , t)\ket{\psi(t)},
\end{align}
where $\ket{\psi(t)} = \hat{U}(t,0)\ket{\psi}$, $\ket{\psi}$ is an initial state vector with arbitrary form, $\tau_i$ represent the relaxation times of different mechanisms, such as those due to electron-electron or electron-phonon interactions, and $\hat{\rho}_i^{\text{eq}}$ are the different density matrices to which the system relaxes for each mechanism. At each time step, the density matrix will contain the complete history of the electron occupation during the evolution, leading to non-thermal electron distributions. 

In section \ref{sec:bilayer} we will study the case of bilayer graphene, where the dominant and fastest relaxation channel is electron–electron scattering. Therefore, we will only consider this mechanism by setting $\tau_i=\tau_\x{ee}$ \cite{Hot_carriers_in_graphene_fundamentals_and_applications_massicotte_giancarlo_soavi_principi_klaas-jan_tielrooij_2021}. The electron-electron relaxation time, $\tau_\x{ee}$, is the average time electrons take to relax to an instantaneous thermal distribution and consequently, $\hat{\rho}_{i}^\x{eq}(t + \Delta t)=\hat{\rho}_{\mathrm{thermal}}(t + \Delta t)$, where at each time step the chemical potential $\mu(t)$ and temperature $T(t)$ are computed to ensure energy and carrier conservation. This relaxation mechanism has been shown to accurately reproduce the optical absorption of graphene \cite{Real_Time_Out_of_Equilibrium_Quantum_Dynamics_in_Disordered_Materials_canonico_roche_cummings_2024}.

From the density matrix at a given time $t$, we compute the conductivity tensor $\sigma_{\alpha\beta}(t)$, which arises when measuring the current along direction $\alpha$, $\hat{J}_{\alpha}$, in response to an applied electric field along direction $\beta$ \cite{Real_Time_Out_of_Equilibrium_Quantum_Dynamics_in_Disordered_Materials_canonico_roche_cummings_2024}. We derive an expression for $\sigma_{\alpha\beta}$ that allows the introduction of a density matrix of arbitrary form by assuming a linear time-dependent perturbation $\hat{V}=-f(t')\hat{\mathbf{r}}\cdot\mathbf{E}$, where $\mathbf{r}$ is the position operator, $\mathbf{E}$ is the applied electric field, and $f(t')$ is the time profile of the perturbation, which is set to turn off adiabatically over a timescale $t_\phi$. The conductivity tensor is then computed as \cite{Nonlinear_optical_susceptibilities_of_semiconductors_Results_with_a_length_gauge_analysis_aversa_sipe_1995, Electrical_Transport_in_Nanoscale_Systems_Di_Ventra_2008, suppmaterial},
\begin{equation}\label{eq:contracted_conductivity}
        \sigma_{\alpha\beta}(t)\!=\!\dfrac{ie}{\hbar}\!\lim_{t_{\phi}\to\infty}\!\int_0^{\infty}\!\!\!\!\!\!\!dt' e^{\!-\frac{t'}{t_{\phi}}}\mathrm{Tr}\left\{\!\left[\hat{r}_{\beta},\hat{\rho}(t)\right]\hat{U}^{\dagger}(t'\!,\!0)\hat{J}_{\alpha}\hat{U}(t'\!,\!0)\!\right\}\!,
\end{equation}
\noindent where $e$ is the electron charge and $\hbar$ is the reduced Planck constant. This formula allows the implementation of different types of currents, such as charge, spin-polarized, or orbital-polarized current. In the following, we will work with the charge current, $\hat{J}_{\alpha} = -e \hat{v}_{\alpha} / \Omega$, where $\Omega$ is the sample area and $\hat{v}_{\alpha}$ is the velocity operator in the $\alpha$ direction. This formula is evaluated using the instantaneous Hamiltonian, $\hat{H}(t)$, and the non-equilibrium density matrix at a given time, which has undergone the time evolution $\hat{\rho}(0) \to \hat{\rho}(t)$. Further details on the derivation of Eq.\ \eqref{eq:contracted_conductivity} are provided in the SM \cite{suppmaterial}.

Additionally, we compute the electron occupation of the excited distribution at $t$ \cite{Real_Time_Out_of_Equilibrium_Quantum_Dynamics_in_Disordered_Materials_canonico_roche_cummings_2024,suppmaterial}, 
\begin{equation}\label{eq:excited_carriers}
    \braket{n(t;E)}=\dfrac{\mathrm{Tr}\left\{\delta\left(\hat{H}(t)-E\right)\hat{\rho}(t)\right\}}{\mathrm{Tr}\left\{\delta\left(\hat{H}(t)-E\right)\right\}},
\end{equation}
where $\delta(\hat{H}-E)$ is the spectral measure operator, i.e., the Dirac delta operator,
which projects the occupation onto energy $E$
\cite{Real_Time_Out_of_Equilibrium_Quantum_Dynamics_in_Disordered_Materials_canonico_roche_cummings_2024}. In sections \ref{sec:haldane} and \ref{sec:bilayer} we will also plot the number of excited carriers multiplied by the density of states (DOS), which is just the numerator of Eq.\ \eqref{eq:excited_carriers}.

\subsection{Real-space numerical implementation}

Because we evolve the density matrix in a vector form, using Eq.\ \eqref{eq:neq_evolution} and then applying Eq.\ \eqref{eq:contracted_conductivity}, we need to open the position and density matrix commutator, leading to 
\begin{gather}
\sigma_{\alpha\beta} (t) = \notag \\
\frac{e^2}{\x{i}\hbar\Omega} \lim_{t_{\phi} \to \infty} 
\int\limits_0^{\infty} \, \x{e}^{-\frac{t'}{t_{\phi}}} \,
 \bra{\rho(t)} \hat{U}^{\dagger}(t',0)\left[\hat{v}_\alpha,\hat{r}_\beta\right] \hat{U}(t',0)\ket{\psi(t)} \notag \\
+ 2\mathrm{Im}\left(\bra{\rho(t)}\hat{U}^{\dagger}(t',0)\hat{v}_\alpha\left[\hat{U}(t',0),\hat{r}_{\beta}\right]\ket{\psi(t)}\right)
dt', \label{eq:neq_sigma}
\end{gather}
where, as we have stated before, $\ket{\rho(t)}$ and $\ket{\psi(t)}$ come from the time evolution of Eq.\ \eqref{eq:neq_evolution}.

In Eqs.\ \eqref{eq:neq_evolution}-\eqref{eq:neq_sigma} we have a number of operators, including $\hat{U}(t_1,t_0)$, $\delta(\hat{H}-E)$, and $\hat{\rho}_{\mathrm{thermal}}$, that are functions of the Hamiltonian. Instead of evaluating these through direct diagonalization, we employ the kernel polynomial method (KPM) and expand them as a series of Chebyshev polynomials \cite{suppmaterial, The_kernel_polynomial_method_weiße_wellein_alvermann_fehske_2006, Linear_scaling_quantum_transport_methodologies_fan_garcia_cummings_barrios-vargas_panhans_harju_ortmann_roche_2020}, $F(\widetilde{H}) \approx \sum_{m=0}^{M}g_m \mu_m \hat{T}_m(\widetilde{H})$, where $g_m$ are the coefficients of the Jackson kernel, used to smooth Gibbs oscillations, $\mu_m$ are the so-called moments of the expansion, and $\hat{T}_m(\widetilde{H})$ is the Chebyshev polynomial of order $m$. Here the Hamiltonian has been normalized such that its energy spectrum lies in $[-1,1]$; $\widetilde{H} = ( \hat{H}-\Bar{E} )/\Delta E$, where $\Bar{E} = (E_{\mathrm{max}}+E_{\mathrm{min}})/2$, $\Delta E = (E_{\mathrm{max}}-E_{\mathrm{min}})/2$, and $E_{\mathrm{min}}$ and $E_{\mathrm{max}}$ are the minimum and maximum eigenvalues of $\hat{H}$, respectively \cite{Linear_scaling_quantum_transport_methodologies_fan_garcia_cummings_barrios-vargas_panhans_harju_ortmann_roche_2020,The_kernel_polynomial_method_weiße_wellein_alvermann_fehske_2006}. The key advantage of the Chebyshev expansion is that it can be evaluated recursively, $\hat{T}_m(\widetilde{H}) = 2 \widetilde{H} \hat{T}_{m-1}(\widetilde{H}) - \hat{T}_{m-2}(\widetilde{H})$, such that the method boils down to a series of matrix-vector multiplies.


Because the position operator is embedded in a commutator, it is well behaved for any type of boundary condition. Finally, the velocity can be easily obtained in the real-space basis as $\bra{r_i}\hat{v}_{\alpha}\ket{r_j} = -i (r_{\alpha;i} - r_{\alpha;j}) H_{ij} / \hbar$, and its commutator with the position operator as $\bra{r_i}[\hat{v}_{\alpha}, \hat{r}_{\beta}]\ket{r_j} = -i (r_{\alpha;i} - r_{\alpha;j})(r_{\beta;i} - r_{\beta;j}) H_{ij} / \hbar$, where $H_{ij}$ are the matrix elements of the Hamiltonian. In practice, we set a finite number of moments $M$, giving rise to a finite energy broadening, $\delta E = \pi \Delta E / M$, which also defines the time of the adiabatic turn-on of the electric field, $t_{\phi} = \hbar / \delta E$ \cite{Linear_scaling_quantum_transport_methodologies_fan_garcia_cummings_barrios-vargas_panhans_harju_ortmann_roche_2020}.

To evaluate the traces in Eqs.\ \eqref{eq:excited_carriers} and \eqref{eq:neq_sigma}, we use the stochastic trace approximation, $\mathrm{Tr}\{\hat{A}\} \approx \frac{1}{R} \sum_{i=1}^R \bra{\psi_i} \hat{A} \ket{\psi_i}   
$where $\ket{\psi_i}$ is a complex vector with random phase at each site and $R$ is the number of random vectors \cite{suppmaterial, Linear_scaling_quantum_transport_methodologies_fan_garcia_cummings_barrios-vargas_panhans_harju_ortmann_roche_2020, The_kernel_polynomial_method_weiße_wellein_alvermann_fehske_2006}. The error of this approximation scales as $1/\sqrt{RN}$ where $N$ is the number of orbitals in the system. We use such random phase states as our initial state in the time evolution of Eq.\ \eqref{eq:neq_evolution}, such that the traces in Eqs.\ \eqref{eq:contracted_conductivity} and \eqref{eq:excited_carriers} become,
\begin{equation}
    \mathrm{Tr}\{ \hat{\rho}(t) \hat{A} \} \approx \frac{1}{R} \sum_{i=1}^R \bra{\psi_i} \hat{\rho}(t) \hat{A} \ket{\psi_i} = \frac{1}{R} \sum_{i=1}^R \bra{\rho_i(t)} \hat{A} \ket{\psi_i},
\end{equation}
\noindent where $\rho_i(t)$ comes from Eq.\ \eqref{eq:neq_evolution} \cite{Real_Time_Out_of_Equilibrium_Quantum_Dynamics_in_Disordered_Materials_canonico_roche_cummings_2024}. As such, the calculated conductivity becomes $\sigma_{\alpha \beta}(t) \approx \frac{1}{R} \sum_{i=1}^{R} \sigma_{\alpha \beta}^i(t)$, where $\sigma_{\alpha \beta}^i(t)$ is one instance of the conductivity calculated from Eq.\ \eqref{eq:neq_sigma} starting from state $\ket{\psi_i}$.

The combination of the Chebyshev polynomial expansion with the stochastic trace approximation makes the methodology $\mathcal{O}(N)$, i.e., the computation time scales linearly with the system size. This allows the computation of systems with millions of atoms -- for the calculations in this work we used between $1 \cdot 10^6$ and $30 \cdot 10^6$ orbitals, which in the latter case corresponds to $2900 \times 2900$ unit cells of bilayer graphene, equivalent to a device with an area of $0.7 \times 0.7$ $\upmu$m$^2$. 

\section{Optical excitation of the Haldane model and gapped graphene}\label{sec:haldane}
\begin{figure*}[tbh]
\includegraphics[width=0.9\textwidth]{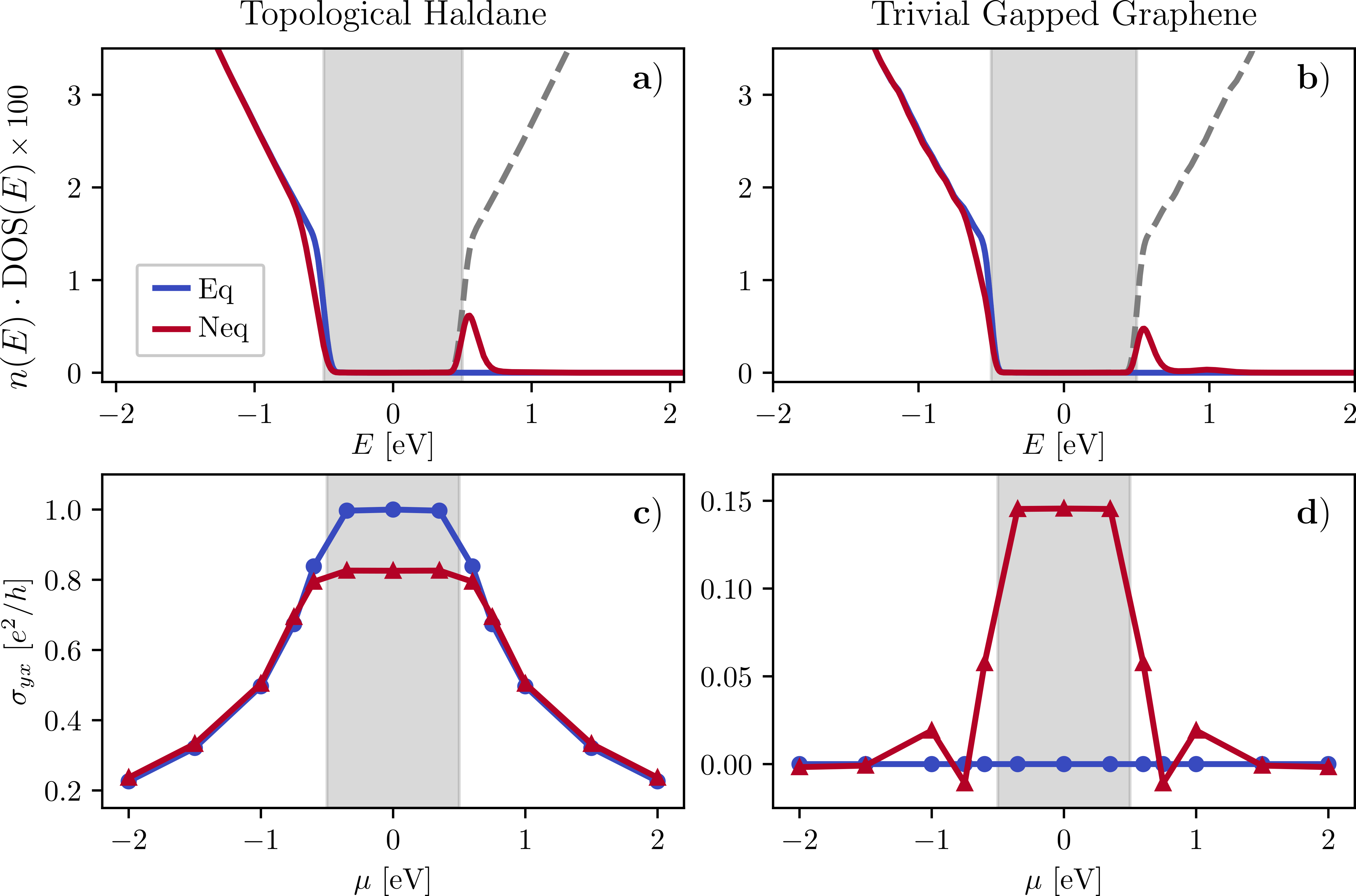}
\caption{a) Number of carriers multiplied by the density of states, at $\mu=0$, for the topological Haldane model. The blue/red solid lines correspond to before and after optical excitation, respectively. The dashed gray line shows the density of states. b) Same as panel a) but for trivially-gapped graphene. c),d) The corresponding Hall conductivities before and after optical excitation. In all panels, the shaded areas represent the band gap. In this figure, all simulations were made in a system with $\sim 1\cdot 10^6$ atoms, $M=1024$, $\hbar\omega=1$ eV and $\Gamma=0.025$.\label{fig:haldane_gap_graphene}}
\end{figure*}
To examine the underlying physical mechanism that drives non-equilibrium topological responses, schematically shown in Fig.\ \ref{fig:optical_scheme}, we introduce the Haldane model \cite{HaldaneModel},
\begin{align}\label{eq:Ham_haldane} 
\hat{H} =& \sum_{\left<ij\right>} \gamma_{ij} a_i^{\dagger} b_j +m\sum_{i}( a_i^{\dagger} a_i-b_i^{\dagger} b_i )\\&+ \sum_{\left<\left<ij\right>\right>} \gamma_{2,ij}( \x{e}^{\x{i}\phi_{ij}}a_i^{\dagger}a_j +\x{e}^{-\x{i}\phi_{ij}}b_i^{\dagger}b_j )+ \mathrm{h.c.}
\end{align}
where $a^\dagger_{i} / a_{i}$ ($b^\dagger_{i} / b_{i}$) are the creation/annihilation operators of an electron at site $i$ in sublattice A (B). The first term describes nearest-neighbor hopping, where we set $\gamma_{ij} = \gamma = -2.7$ eV, the value in graphene. The second term, positive (negative) on sublattice $A$ ($B$), breaks inversion symmetry and opens a trivial band gap $\Delta_m = 2m$. The third term is a complex second-neighbor hopping that breaks time-reversal symmetry and opens a topological gap $\Delta_\mathrm{H} = 6\sqrt{3}|\gamma_2|$ (here we set $\phi_{ij}=\pi/2$). When both gap-opening terms are nonzero, the total band gap is $\Delta = |\Delta_m - \Delta_{\mathrm{H}}|$. When $\Delta_{m} > \Delta_{\mathrm{H}}$ the system is a trivial insulator, otherwise it is a Chern insulator \cite{HaldaneModel}. In equilibrium, the Hall conductivity within the band gap is given by $\sigma_{yx}=\mathcal{C} \cdot e^2/h$, where $\mathcal{C}=0$ $(1)$ is the Chern number in the trivial (topological) phase \cite{HaldaneModel}.

We consider the topological phase by setting $m=0$ and $\gamma_2=1/(6\sqrt{3})$ eV, and the trivial phase by setting $m=0.5$ eV and $\gamma_2=0$. Both phases have the same DOS with a band gap of $1$ eV, as indicated by the dashed gray lines in Figs.\ \ref{fig:haldane_gap_graphene}(a,b). Despite their identical DOS, the difference between these two systems is revealed in their equilibrium Hall conductivity, calculated using Eq.\ \eqref{eq:neq_sigma} at $t=0$ and shown as the blue lines in Figs.\ \ref{fig:haldane_gap_graphene}(c,d). As mentioned above, the topological phase exhibits $\sigma_{yx} = e^2/h$ within the gap, while for the trivial phase $\sigma_{yx} = 0$.

This behavior can be understood by considering the Chern number, which arises from the sum of the Berry curvature, $\Omega_{yx}$, in both valleys. Depending on the Berry curvature polarization in each valley, the sum will result in $\mathcal{C} = 0$ or $1$. In the topological case, $\Omega_{yx} > 0$ in both valleys, summing up to a non-zero contribution. Meanwhile, in the trivial phase, $\Omega_{yx}$ has opposite sign in each valley, perfectly canceling and leading to $\mathcal{C} = 0$ \cite{HaldaneModel}. In addition, the Berry curvature has opposite sign in the valence and conduction bands, resulting in a vanishing Hall conductivity outside of the gap.

To drive the system out of equilibrium, we apply an optical pulse with vector potential $\mathbf{A}(t) = A_0 P(t) \left(\cos(\omega_\x{p} t)\hat{x} + \eta\sin(\omega_\x{p} t)\hat{y}\right)$, where $A_0$ is the amplitude, $P(t)$ is the pulse envelope, $\hbar \omega_\x{p}$ is the photon energy, and $\eta = \pm 1$ corresponds to right/left circular polarization \cite{Tuning_laser_induced_band_gaps_in_graphene_calvo_pastawski_roche_luis_2011}. The envelope function is set as $P(t) = \mathrm{sech}[(t-2T_\x{p})/(b T_\x{p})]$. Here we consider a short pulse of just a few optical periods by letting $T_\x{p} = 2\pi/\omega_\x{p}$ and $b \approx 0.5673$, such that the full width at half maximum of $P^2(t)$ is $T_\x{p}$ \cite{Ultrafast_Lasers_keller_2021}. Using the Peierls substitution, the neighbor hoppings in the Hamiltonian become time dependent,
\begin{equation} \gamma_{ij}(t) = \gamma \exp\left(\x{i}\dfrac{2\pi}{\Phi_0}\int_{\mathrm{r}_i}^{\mathrm{r}_j}d\mathbf{r}\cdot\mathbf{A}(t)\right), \end{equation}
where $\mathbf{r}_i$ is the position of site $i$ and $\Phi_0=h/e$ is the magnetic flux quantum. In the following, we set $A_0=\Gamma\Phi_0/2a$, where $\Gamma$ is a dimensionless free parameter that controls the strength of the pulse and $a=0.246$ nm is the graphene lattice constant.

In what follows, we consider irradiation times up to $4T_\x{p}$, covering the full range of the optical pulse. During this period, we evolve the system using Eq.\ \eqref{eq:neq_evolution}, and once the pulse ends we calculate the conductivity and the number of excited carriers of the resulting out-of-equilibrium system using Eqs.\ \eqref{eq:neq_sigma} and \eqref{eq:excited_carriers}. We consider a circularly-polarized optical pulse with a photon energy of $\hbar\omega_\x{p} = 1$ eV, corresponding to the topological or trivial band gap, and a pulse strength of $\Gamma=2.5 \cdot 10^{-2}$. At this photon energy, $4T_\x{p} \approx 16$ fs, shorter than the typical electron-electron thermalization time of graphene, so for these simulations we let $\tau_{ee} \to \infty$.


The solid lines in Fig.\ \ref{fig:haldane_gap_graphene} a) show the carrier occupation before (blue) and after (red) optical excitation, indicating a transfer of carriers from $-\hbar\omega_\x{p}/2$ to $+\hbar\omega_\x{p}/2$ for both the topological and the trivial phases. The higher number of transferred carriers in the topological case reflects the fact that both valleys are excited, while in the trivial case only one valley is excited (see Fig.\ \ref{fig:optical_scheme}).
These changes in carrier occupation manifest in the Hall conductivity, $\sigma_{yx}$, which from this point onward is defined as the difference between left and right circular polarization, $\sigma_{yx} = (\sigma_{yx}^{\circlearrowleft} - \sigma_{yx}^{\circlearrowright})/2$, following the convention commonly used in experiments \cite{Light_induced_anomalous_Hall_effect_in_graphene_mciver_schulte_stein_matsuyama_jotzu_meier_cavalleri_2019}. This behavior is illustrated in Figs.\ \ref{fig:haldane_gap_graphene}(c,d).

In the topological phase at $\mu=0$, circularly polarized light excites all available carriers in both valleys, transferring electrons from regions with positive Berry curvature to regions with negative Berry curvature \cite{Perfect_Circular_Dichroism_in_the_Haldane_Model_kazu_ghalamkari_tatsumi_saito_2018}. As a result, the total Hall conductivity becomes smaller than in the equilibrium case.
In the trivial phase, circularly polarized light excites electrons from only one valley \cite{Perfect_Circular_Dichroism_in_the_Haldane_Model_kazu_ghalamkari_tatsumi_saito_2018}, which now contributes with the opposite sign of Berry curvature. As the contributions from the two valleys no longer cancel, this leads to a non-zero Hall conductivity.
In both cases, the resulting non-equilibrium conductivity can be tuned by adjusting the pulse duration or intensity, which controls the number of excited carriers and, consequently, the total integral of the Berry curvature over the occupied states. 

As we will see in section \ref{sec:time_bernal}, the total value of the Hall conductivity in the out-of-equilibrium regime can even exceed the expected case of one valley being fully empty and the other fully occupied. This is due to the fact that, in the out-of-equilibrium regime, the conductivity is not determined solely by the Chern number of the bands, but also contains additional terms that emerge from the off-diagonal elements of the density matrix during its time evolution.

\section{Carrier thermalization and optically excited bilayer graphene}\label{sec:bilayer}

\begin{figure}[tbh]
\includegraphics[width=0.95\columnwidth]{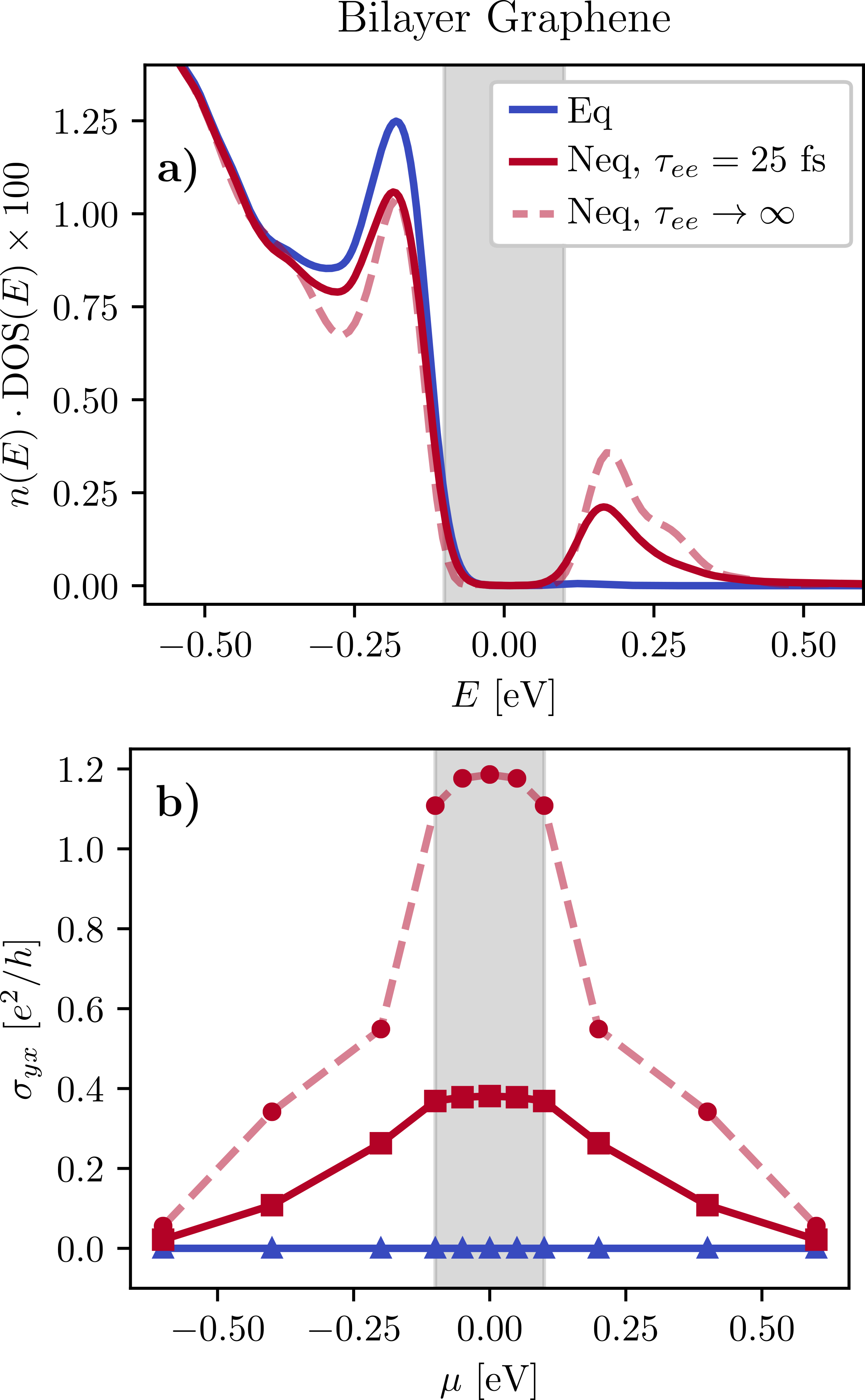}
\caption{a) Number of carriers multiplied by the density of states, at $\mu=0$, for gapped bilayer graphene. The blue line represents the equilibrium case, and the red solid (dashed) line shows the case after optical excitation for carrier thermalization time $\tau_\x{ee}= 25$ fs ($\tau_\x{ee} \to \infty$).
b) The corresponding Hall conductivity as a function of the chemical potential. In this figure, all simulations were performed in a system with $\sim 30\cdot 10^6$ atoms, $M=2048$, $\hbar\omega=0.2$ eV, and $\Gamma=0.025$, which corresponds to a fluence of $F=0.04$ $\mathrm{mJ/cm^2}$.}\label{fig:bernal}
\end{figure}

\begin{figure*}[tbh]
\includegraphics[width=0.85\textwidth]{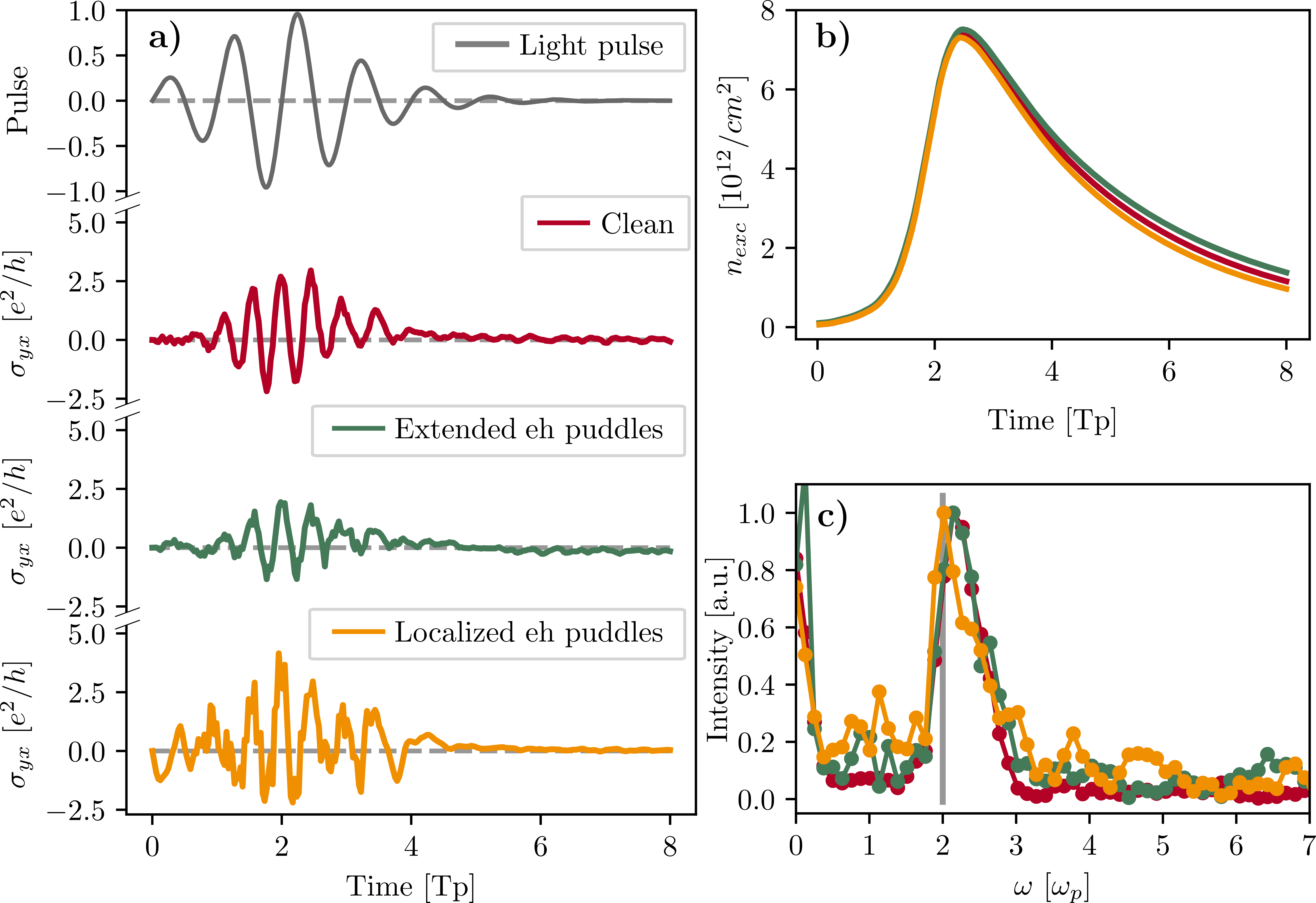}
\caption{a) Time dependence of the applied light pulse in gray, and the Hall conductivity in red, green, and yellow for the clean system, extended electron-hole puddles, and localized electron-hole puddles, respectively, at $\mu = 0$.
b) Number of excited carriers for the different systems.
c) Fourier transform of the time-dependent conductivity, with the frequency in units of the pulse frequency. In this figure, all simulations were performed in a system with $\sim 30\cdot 10^6$ atoms, $M=2048$, $\hbar\omega=0.2$ eV, $\tau_{ee}=25$ fs, and $\Gamma=0.025$, which corresponds to a fluence of $F=0.04$ $\mathrm{mJ/cm^2}$.\label{fig:sigma_time}}
\end{figure*}

Next we examine the non-equilibrium topological response in a more experimentally accessible system, AB-stacked (Bernal) bilayer graphene (BLG), which can be modeled using the tight-binding Hamiltonian \cite{The_electronic_properties_of_bilayer_graphene_mccann_koshino_2013, Spin_dynamics_in_bilayer_graphene_Role_of_electron-hole_puddles_and_Dyakonov-Perel_mechanism_tuan_adam_roche_2016}, 
\begin{equation}\label{eq:Ham_bernal} 
\begin{aligned}
    \hat{H} =& \sum_{l=-1,1} \sum_{\left<ij\right>} \gamma_{ij} a_{l,i}^{\dagger} b_{l,j} +\gamma_{1}\sum_{i}  a_{-1,i}^{\dagger} b_{1,i} \\
    &+\frac{U}{2} \sum_{l=-1,1} l \sum_{i} \left( a_{l,i}^{\dagger}a_{l,i}+b_{l,i}^{\dagger}b_{l,i} \right)
    +\x{h.c.}
\end{aligned}
\end{equation}
\noindent Here the subindex $l$ in the creation/annihilation operators refers to the layer, indexed by $l = \pm 1$.
The first term in Eq.\ \eqref{eq:Ham_bernal} describes the intralayer nearest-neighbor hopping; we set $\gamma_{ij} = \gamma_0 = -2.7$ eV. The second term is an interlayer coupling that connects dimer sites between layers; we set $\gamma_1 = 0.4$ eV. The last term breaks interlayer symmetry and consequently opens a band gap \cite{The_electronic_properties_of_bilayer_graphene_mccann_koshino_2013}. This can be induced experimentally through doping or the use of external gates. In the following, we set $U = 0.23$ eV, which opens a gap of $\Delta = 0.2$ eV, a value shown to be experimentally realizable \cite{Direct_observation_of_a_widely_tunable_bandgap_in_bilayer_graphene_zhang_tang_girit_hao_martin_zettl_crommie_shen_wang_2009}.
In equilibrium, the system exhibits zero Hall conductivity, which occurs because both valleys have opposite Berry curvature \cite{The_electronic_properties_of_bilayer_graphene_mccann_koshino_2013}, leading to cancellation of the total topological response.

In this section we also consider a circularly-polarized optical pulse, but with a photon energy of $\hbar\omega_\x{p} = 0.2$ eV, corresponding to the band gap, and a pulse strength of $\Gamma=2.5 \cdot 10^{-2}$, which corresponds to a fluence of $F=0.04$ $\mathrm{mJ/cm^2}$ \cite{suppmaterial}. At this photon energy, the total simulation time is $4T_\x{p} \approx 83$ fs, and thus the electron-electron relaxation time in this system, $\tau_\x{ee} = 25$ fs \cite{Hot_carriers_in_graphene_fundamentals_and_applications_massicotte_giancarlo_soavi_principi_klaas-jan_tielrooij_2021, Real_Time_Out_of_Equilibrium_Quantum_Dynamics_in_Disordered_Materials_canonico_roche_cummings_2024}, will play a relevant role in the electron dynamics. 

In the blue curve of Fig.\ \ref{fig:bernal}a), we show the equilibrium DOS multiplied by the electron occupation number $\braket{n(0;E)}\cdot\mathrm{DOS}(E)$, at $\mu=0$ eV prior to optical excitation. When the system is driven out of equilibrium, the optical pulse promotes carriers from $-\hbar\omega_\x{p}/2$ to $\hbar\omega_\x{p}/2$, as shown by the red curves, where the peaks are shifted due to the convolution of the excited carriers with the shape of the DOS. In the limit $\tau_\x{ee} \to \infty$, all the terms with $\tau_i$ that appear in Eq.\ \ref{eq:neq_evolution} go to zero and the time evolution doesn't exhibit any energy relaxation. This limit is shown by the dashed red curve in Fig.\ \ref{fig:bernal}a), where we can also see the excitation of carriers to higher energies. When a finite thermalization time is introduced (solid red curve), this second peak smoothens, as the excited carriers now thermalize into a hot electron distribution \cite{Hot_carriers_in_graphene_fundamentals_and_applications_massicotte_giancarlo_soavi_principi_klaas-jan_tielrooij_2021}.

These changes in carrier occupation manifest in the Hall conductivity at the end of the light pulse, $\sigma_{yx}(t=4T_\x{p})$, illustrated by the red curves of Fig.\ \ref{fig:bernal}b) for different chemical potentials. As in section \ref{sec:haldane}, circularly polarized light excites electrons from only one valley \cite{Perfect_Circular_Dichroism_in_the_Haldane_Model_kazu_ghalamkari_tatsumi_saito_2018} leading to a non-zero Hall conductivity, in agreement with the trends observed in experiments \cite{Tunable_and_giant_valley_selective_Hall_effect_in_gapped_bilayer_graphene_yin_tan_barcons-ruiz_torre_watanabe_taniguchi_song_hone_koppens_2022}. Introducing a finite carrier thermalization time significantly suppresses the response. This reduction occurs because carrier relaxation tends to drive the excited carrier population toward a hot-electron distribution that mainly occupies the valence band. As for the case of the Haldane model and gapped single-layer graphene, the value of the non-equilibrium conductivity can be tuned by adjusting the pulse duration or intensity, which controls the number of excited carriers.


\section{Dynamics of the light-induced Hall conductivity in disordered BLG}\label{sec:time_bernal}

We now proceed to study the intertwined role of energy transfer dynamics and disorder effects on the optically-driven Hall conductivity. In Fig.\ \ref{fig:sigma_time}a), we show the time evolution of the out-of-equilibrium Hall conductivity over a time period $8T_\x{p} \approx 166\ \text{fs}$, where the pulse is shown in gray.

As a baseline, we consider the clean case at $\mu = 0\ \text{eV}$, shown by the dark red curve. As can be seen, the out-of-equilibrium conductivity emerges with the onset of the pulse and oscillates between positive and negative values. This change in sign is due to the oscillation of the effective magnetic field induced by the circularly polarized light. The main frequency of oscillation, as shown by the Fourier transform of the Hall signal in Fig.\ \ref{fig:sigma_time}c), is twice that of the applied optical pulse, $2\omega_\x{p}$. This behavior is a special case of the so-called heterodyne effect \cite{Heterodyne_Hall_effect_in_a_two_dimensional_electron_gas}, which manifests as a Hall response arising from the oscillating effective magnetic field of the optical pulse. Specifically, this arises from the combination of time-translation symmetry and a rotation in reciprocal space \cite{Time-resolved_Hall_conductivity_of_pulse_driven_topological_quantum_systems}, and can also be observed in gapped graphene \cite{Time-resolved_Hall_conductivity_of_pulse_driven_topological_quantum_systems,suppmaterial}. The fact that the total magnitude of the Hall conductivity is higher than just the summation of the Berry curvatures when one valley is completely empty and the other completely full is because, in this far-from-equilibrium regime, the conductivity is no longer only the sum of the Berry curvatures and has additional terms that arise from the off-diagonal elements of the density matrix \cite{Time-resolved_Hall_conductivity_of_pulse_driven_topological_quantum_systems,Generation_of_Ultrahigh_Anomalous_Hall_Conductivities_via_Optimally_Prepared_Topological_Floquet_States}.

When the laser pulse ends, the conductivity exponentially decays toward its equilibrium value on the time scale of the carrier-carrier scattering time, $\tau_\x{ee}$. This behavior can also be observed in the number of excited carriers shown in Fig.\ \ref{fig:sigma_time}b). Increasing the carrier–carrier scattering time induces a slower decay toward the equilibrium regime, and in the limit $\tau_\x{ee} \to \infty$, this can lead to nonzero persistent values of the Hall conductivity \cite{suppmaterial}.

By comparing Figs.\ \ref{fig:sigma_time}a) and b), one can see that not only does the number of excited carriers affect the final value of the conductivity, but also their nature and distribution. At two distinct times where the number of excited carriers is the same, $t = 2T_\x{p}$ and $t = 4T_\x{p}$, the Hall conductivity is completely different: the former corresponds to the peak value of $\sigma_{yx}$, while at the latter it is nearly zero. This effect arises because, at the beginning of the pulse, the excited carriers are concentrated at the band edge where the Berry curvature is large, leading to a larger Hall conductivity. At later times, the electrons have thermalized into a broader range of energies within the conduction band, resulting in a reduced net Berry curvature contribution and, therefore, a smaller Hall conductivity.

Next, we consider the presence of disorder in the form of electron-hole puddles. Arising from trapped charges in the substrate, electron-hole puddles are a type of long-range disorder that is ubiquitous in graphene devices \cite{Observation_of_electron_hole_puddles_in_graphene_using_a_scanning_single_electron_transistor_martin_akerman_ulbricht_lohmann_smet_von_klitzing_yacoby_2007, Spatially_resolved_spectroscopy_of_monolayer_graphene_onSiO2_deshpande_bao_miao_lau_leroy_2009} and may be modeled as a series of Gaussian electrostatic impurities added to the onsite energy of the graphene layer \cite{Mechanism_for_puddle_formation_in_graphene_adam_jung_klimov_zhitenev_stroscio_stiles_2011, Spin_dynamics_in_bilayer_graphene_Role_of_electron-hole_puddles_and_Dyakonov-Perel_mechanism_tuan_adam_roche_2016}, $V(\mathbf{r}_i) = \sum_{j=1}^{N_\x{eh}} W_j \exp\left[-\frac{1}{2}\left(\frac{\mathbf{r}_i - \mathbf{R}_j}{\xi}\right)^2\right]$, where $\mathbf{r}_i$ is the position of each carbon atom, $\mathbf{R}_j$ is the center of each puddle, $N_\x{eh}$ is the number of puddles, $\xi$ is the puddle width, and $W_j$ is chosen randomly in the interval $[-W, W]$. We consider two different sets of parameters. The first set corresponds to spatially-extended puddles that appear in graphene on an hBN substrate, where we set $W = 35$ meV and $\xi = 3.5$ nm with a concentration $n = N_\x{eh}/N = 0.0004$ \cite{Microscopic_polarization_in_bilayer_graphene_rutter_jung_klimov_newell_zhitenev_stroscio_2011, Mechanism_for_puddle_formation_in_graphene_adam_jung_klimov_zhitenev_stroscio_stiles_2011}, which leads mainly to intravalley scattering. Another stronger and more localized puddle configuration, with $W = 500$ meV and $\xi = 0.1$ nm and a concentration $n = N_\x{eh}/N = 0.001$, gives rise to both intra- and intervalley scattering processes.

In the case of extended electron-hole puddles, the conductivity remains similar to the clean case, but with a $\sim$$30\%$ reduction in magnitude arising from intravalley scattering, as shown by the green curve in Fig.\ \ref{fig:sigma_time}a). In contrast, the presence of strong localized puddles leads to an increase in the Hall conductivity fluctuations. This point requires further examination, but may arise from the onset of extrinsic effects, such as skew scattering, induced by the localized electrostatic impurities, which may enhance the extra terms that appear in the out-of-equilibrium conductivity and are not directly related to the Chern number.
In both configurations, $2\omega_\x{p}$ remains the dominant component in the frequency spectrum, although we observe that additional frequencies appear in the disordered cases, particularly for the strong localized puddles. This is due to the  disorder-induced breaking of the reciprocal space symmetry, previously presented in the clean case.


\section{Summary and conclusions}
We have introduced a linear-scaling, real-space quantum transport framework for nonperturbative simulations of driven disordered quantum materials deep in the far-from-equilibrium regime. Applying this methodology, we demonstrate that ultrafast circularly polarized optical pulses can dynamically generate (suppress) a finite Hall response in an otherwise topologically trivial (non-trivial) two-dimensional material under realistic conditions. The Hall conductivity emerges in real time during driving and oscillates at twice the optical frequency, reflecting the heterodyne character of the optically-induced effective magnetic field. The Hall response then decays after the optical pulse as thermalization restores equilibrium, opening the path for the design of ultrafast optical detectors. This transient Hall response is robust against long-range electrostatic disorder and can even be enhanced by short-range inhomogeneities. This enhancement may arise from the activation of extrinsic Hall effects by strong localized disorder. More broadly, the generality of our framework and its linear‑scaling $\mathcal{O}(N)$ character make it well-suited for integration with emerging AI-assisted Hamiltonian design and discovery approaches, providing a scalable pathway to engineer and explore light-induced topological and transport phenomena in complex disordered materials and heterostructures, where nonequilibrium energy flow and disorder may enable qualitatively new regimes of functionality.

\begin{acknowledgments}
We acknowledge funding from MCIN/AEI /10.13039/501100011033 and European Union ”NextGenerationEU/PRTR” under grant PCI2021-122035-2A-2a. ICN2 is funded by the CERCA Programme Generalitat de Catalunya and supported by the Severo Ochoa Centres of Excellence programme, Grant CEX2021-001214-S, funded
by MCIN/AEI/10.13039.501100011033. This work is also supported by MICIN with European funds-NextGenerationEU (PRTR-C17.I1) and by and 2021 SGR 00997, funded by Generalitat de Catalunya.
Work performed at the Center for Nanoscale Materials, a U.S. Department of Energy Office of Science User Facility, was supported by the U.S. DOE, Office of Basic Energy Sciences, under Contract No. DE-AC02-06CH11357.
\end{acknowledgments}


\begin{thebibliography}{69}%
\makeatletter
\providecommand \@ifxundefined [1]{%
 \@ifx{#1\undefined}
}%
\providecommand \@ifnum [1]{%
 \ifnum #1\expandafter \@firstoftwo
 \else \expandafter \@secondoftwo
 \fi
}%
\providecommand \@ifx [1]{%
 \ifx #1\expandafter \@firstoftwo
 \else \expandafter \@secondoftwo
 \fi
}%
\providecommand \natexlab [1]{#1}%
\providecommand \enquote  [1]{``#1''}%
\providecommand \bibnamefont  [1]{#1}%
\providecommand \bibfnamefont [1]{#1}%
\providecommand \citenamefont [1]{#1}%
\providecommand \href@noop [0]{\@secondoftwo}%
\providecommand \href [0]{\begingroup \@sanitize@url \@href}%
\providecommand \@href[1]{\@@startlink{#1}\@@href}%
\providecommand \@@href[1]{\endgroup#1\@@endlink}%
\providecommand \@sanitize@url [0]{\catcode `\\12\catcode `\$12\catcode `\&12\catcode `\#12\catcode `\^12\catcode `\_12\catcode `\%12\relax}%
\providecommand \@@startlink[1]{}%
\providecommand \@@endlink[0]{}%
\providecommand \url  [0]{\begingroup\@sanitize@url \@url }%
\providecommand \@url [1]{\endgroup\@href {#1}{\urlprefix }}%
\providecommand \urlprefix  [0]{URL }%
\providecommand \Eprint [0]{\href }%
\providecommand \doibase [0]{https://doi.org/}%
\providecommand \selectlanguage [0]{\@gobble}%
\providecommand \bibinfo  [0]{\@secondoftwo}%
\providecommand \bibfield  [0]{\@secondoftwo}%
\providecommand \translation [1]{[#1]}%
\providecommand \BibitemOpen [0]{}%
\providecommand \bibitemStop [0]{}%
\providecommand \bibitemNoStop [0]{.\EOS\space}%
\providecommand \EOS [0]{\spacefactor3000\relax}%
\providecommand \BibitemShut  [1]{\csname bibitem#1\endcsname}%
\let\auto@bib@innerbib\@empty
\bibitem [{\citenamefont {Kobayashi}\ \emph {et~al.}(2023)\citenamefont {Kobayashi}, \citenamefont {Heide}, \citenamefont {Johnson}, \citenamefont {Tiwari}, \citenamefont {Liu}, \citenamefont {Reis}, \citenamefont {Heinz},\ and\ \citenamefont {Ghimire}}]{Floquet_engineering_of_strongly_driven_excitons_in_monolayer_tungsten_disulfide}%
  \BibitemOpen
  \bibfield  {author} {\bibinfo {author} {\bibfnamefont {Y.}~\bibnamefont {Kobayashi}}, \bibinfo {author} {\bibfnamefont {C.}~\bibnamefont {Heide}}, \bibinfo {author} {\bibfnamefont {A.~C.}\ \bibnamefont {Johnson}}, \bibinfo {author} {\bibfnamefont {V.}~\bibnamefont {Tiwari}}, \bibinfo {author} {\bibfnamefont {F.}~\bibnamefont {Liu}}, \bibinfo {author} {\bibfnamefont {D.~A.}\ \bibnamefont {Reis}}, \bibinfo {author} {\bibfnamefont {T.~F.}\ \bibnamefont {Heinz}},\ and\ \bibinfo {author} {\bibfnamefont {S.}~\bibnamefont {Ghimire}},\ }\bibfield  {title} {\bibinfo {title} {Floquet engineering of strongly driven excitons in monolayer tungsten disulfide},\ }\href {https://doi.org/10.1038/s41567-022-01849-9} {\bibfield  {journal} {\bibinfo  {journal} {Nature Physics}\ }\textbf {\bibinfo {volume} {19}},\ \bibinfo {pages} {171} (\bibinfo {year} {2023})}\BibitemShut {NoStop}%
\bibitem [{\citenamefont {Mitrano}\ \emph {et~al.}(2016)\citenamefont {Mitrano}, \citenamefont {Cantaluppi}, \citenamefont {Nicoletti}, \citenamefont {Kaiser}, \citenamefont {Perucchi}, \citenamefont {Lupi}, \citenamefont {Pietro}, \citenamefont {Pontiroli}, \citenamefont {Riccò}, \citenamefont {Clark}, \citenamefont {Jaksch},\ and\ \citenamefont {Cavalleri}}]{Possible_light_induced_superconductivity_in_K3C60_at_high_temperature}%
  \BibitemOpen
  \bibfield  {author} {\bibinfo {author} {\bibfnamefont {M.}~\bibnamefont {Mitrano}}, \bibinfo {author} {\bibfnamefont {A.}~\bibnamefont {Cantaluppi}}, \bibinfo {author} {\bibfnamefont {D.}~\bibnamefont {Nicoletti}}, \bibinfo {author} {\bibfnamefont {S.}~\bibnamefont {Kaiser}}, \bibinfo {author} {\bibfnamefont {A.}~\bibnamefont {Perucchi}}, \bibinfo {author} {\bibfnamefont {S.}~\bibnamefont {Lupi}}, \bibinfo {author} {\bibfnamefont {P.~D.}\ \bibnamefont {Pietro}}, \bibinfo {author} {\bibfnamefont {D.}~\bibnamefont {Pontiroli}}, \bibinfo {author} {\bibfnamefont {M.}~\bibnamefont {Riccò}}, \bibinfo {author} {\bibfnamefont {S.~R.}\ \bibnamefont {Clark}}, \bibinfo {author} {\bibfnamefont {D.}~\bibnamefont {Jaksch}},\ and\ \bibinfo {author} {\bibfnamefont {A.}~\bibnamefont {Cavalleri}},\ }\bibfield  {title} {\bibinfo {title} {Possible light-induced superconductivity in k$_3$c$_{60}$ at high temperature},\ }\href {https://doi.org/10.1038/nature16522} {\bibfield  {journal} {\bibinfo  {journal} {Nature}\ }\textbf
  {\bibinfo {volume} {530}},\ \bibinfo {pages} {461} (\bibinfo {year} {2016})}\BibitemShut {NoStop}%
\bibitem [{\citenamefont {Bloch}\ \emph {et~al.}(2022)\citenamefont {Bloch}, \citenamefont {Cavalleri}, \citenamefont {Galitski}, \citenamefont {Hafezi},\ and\ \citenamefont {Rubio}}]{Strongly_correlated_electron--photon_systems}%
  \BibitemOpen
  \bibfield  {author} {\bibinfo {author} {\bibfnamefont {J.}~\bibnamefont {Bloch}}, \bibinfo {author} {\bibfnamefont {A.}~\bibnamefont {Cavalleri}}, \bibinfo {author} {\bibfnamefont {V.}~\bibnamefont {Galitski}}, \bibinfo {author} {\bibfnamefont {M.}~\bibnamefont {Hafezi}},\ and\ \bibinfo {author} {\bibfnamefont {A.}~\bibnamefont {Rubio}},\ }\bibfield  {title} {\bibinfo {title} {Strongly correlated electron--photon systems},\ }\href {https://doi.org/10.1038/s41586-022-04726-w} {\bibfield  {journal} {\bibinfo  {journal} {Nature}\ }\textbf {\bibinfo {volume} {606}},\ \bibinfo {pages} {41} (\bibinfo {year} {2022})}\BibitemShut {NoStop}%
\bibitem [{\citenamefont {Hedwig}\ \emph {et~al.}(2025)\citenamefont {Hedwig}, \citenamefont {Zinke}, \citenamefont {Braun}, \citenamefont {Arnoldi}, \citenamefont {Pulkkinen}, \citenamefont {Min\'ar}, \citenamefont {Ebert}, \citenamefont {Aeschlimann},\ and\ \citenamefont {Stadtm\"uller}}]{Transition_from_Optically_Excited_to_Intrinsic_Spin_Polarization_in_WSe_2}%
  \BibitemOpen
  \bibfield  {author} {\bibinfo {author} {\bibfnamefont {S.}~\bibnamefont {Hedwig}}, \bibinfo {author} {\bibfnamefont {G.}~\bibnamefont {Zinke}}, \bibinfo {author} {\bibfnamefont {J.}~\bibnamefont {Braun}}, \bibinfo {author} {\bibfnamefont {B.}~\bibnamefont {Arnoldi}}, \bibinfo {author} {\bibfnamefont {A.}~\bibnamefont {Pulkkinen}}, \bibinfo {author} {\bibfnamefont {J.}~\bibnamefont {Min\'ar}}, \bibinfo {author} {\bibfnamefont {H.}~\bibnamefont {Ebert}}, \bibinfo {author} {\bibfnamefont {M.}~\bibnamefont {Aeschlimann}},\ and\ \bibinfo {author} {\bibfnamefont {B.}~\bibnamefont {Stadtm\"uller}},\ }\bibfield  {title} {\bibinfo {title} {Transition from optically excited to intrinsic spin polarization in ${\mathrm{wse}}_{2}$},\ }\href {https://doi.org/10.1103/lc2p-kl4y} {\bibfield  {journal} {\bibinfo  {journal} {Phys. Rev. Lett.}\ }\textbf {\bibinfo {volume} {135}},\ \bibinfo {pages} {186903} (\bibinfo {year} {2025})}\BibitemShut {NoStop}%
\bibitem [{\citenamefont {Higuchi}\ \emph {et~al.}(2017)\citenamefont {Higuchi}, \citenamefont {Heide}, \citenamefont {Ullmann}, \citenamefont {Weber},\ and\ \citenamefont {Hommelhoff}}]{Light-field-driven_currents_in_graphene_higuchi_heide_ullmann_weber_hommelhoff_2017}%
  \BibitemOpen
  \bibfield  {author} {\bibinfo {author} {\bibfnamefont {T.}~\bibnamefont {Higuchi}}, \bibinfo {author} {\bibfnamefont {C.}~\bibnamefont {Heide}}, \bibinfo {author} {\bibfnamefont {K.}~\bibnamefont {Ullmann}}, \bibinfo {author} {\bibfnamefont {H.~B.}\ \bibnamefont {Weber}},\ and\ \bibinfo {author} {\bibfnamefont {P.}~\bibnamefont {Hommelhoff}},\ }\bibfield  {title} {\bibinfo {title} {Light-field-driven currents in graphene},\ }\href {https://doi.org/https://doi.org/10.1038/nature23900} {\bibfield  {journal} {\bibinfo  {journal} {Nature}\ }\textbf {\bibinfo {volume} {550}},\ \bibinfo {pages} {224–228} (\bibinfo {year} {2017})}\BibitemShut {NoStop}%
\bibitem [{\citenamefont {Bao}\ \emph {et~al.}(2021)\citenamefont {Bao}, \citenamefont {Tang}, \citenamefont {Sun},\ and\ \citenamefont {Zhou}}]{Light_induced_emergent_phenomena_in_2D_materials_and_topological_materials_bao_tang_sun_zhou_2021}%
  \BibitemOpen
  \bibfield  {author} {\bibinfo {author} {\bibfnamefont {C.}~\bibnamefont {Bao}}, \bibinfo {author} {\bibfnamefont {P.}~\bibnamefont {Tang}}, \bibinfo {author} {\bibfnamefont {D.}~\bibnamefont {Sun}},\ and\ \bibinfo {author} {\bibfnamefont {S.}~\bibnamefont {Zhou}},\ }\bibfield  {title} {\bibinfo {title} {Light-induced emergent phenomena in {2D} materials and topological materials},\ }\href {https://doi.org/10.1038/s42254-021-00388-1} {\bibfield  {journal} {\bibinfo  {journal} {Nature Reviews Physics}\ }\textbf {\bibinfo {volume} {4}},\ \bibinfo {pages} {33–48} (\bibinfo {year} {2021})}\BibitemShut {NoStop}%
\bibitem [{\citenamefont {Rudner}\ and\ \citenamefont {Lindner}(2020)}]{Band_structure_engineering_and_non_equilibrium_dynamics_in_Floquet_topological_insulators_rudner_lindner_2020}%
  \BibitemOpen
  \bibfield  {author} {\bibinfo {author} {\bibfnamefont {M.~S.}\ \bibnamefont {Rudner}}\ and\ \bibinfo {author} {\bibfnamefont {N.~H.}\ \bibnamefont {Lindner}},\ }\bibfield  {title} {\bibinfo {title} {Band structure engineering and non-equilibrium dynamics in {Floquet} topological insulators},\ }\href {https://doi.org/10.1038/s42254-020-0170-z} {\bibfield  {journal} {\bibinfo  {journal} {Nature Reviews Physics}\ }\textbf {\bibinfo {volume} {2}},\ \bibinfo {pages} {229–244} (\bibinfo {year} {2020})}\BibitemShut {NoStop}%
\bibitem [{\citenamefont {Caruso}\ \emph {et~al.}(2025)\citenamefont {Caruso}, \citenamefont {Sentef}, \citenamefont {Attaccalite}, \citenamefont {Bonitz}, \citenamefont {Draxl}, \citenamefont {De~Giovannini}, \citenamefont {Eckstein}, \citenamefont {Ernstorfer}, \citenamefont {Fechner}, \citenamefont {Grüning}, \citenamefont {Hübener}, \citenamefont {Joost}, \citenamefont {Juraschek}, \citenamefont {Karrasch}, \citenamefont {Kennes}, \citenamefont {Latini}, \citenamefont {Lu}, \citenamefont {Neufeld}, \citenamefont {Perfetto}, \citenamefont {Rettig}, \citenamefont {Pela}, \citenamefont {Rubio}, \citenamefont {Rudzinski}, \citenamefont {Ruggenthaler}, \citenamefont {Sangalli}, \citenamefont {Schüler}, \citenamefont {Shallcross}, \citenamefont {Sharma}, \citenamefont {Stefanucci},\ and\ \citenamefont {Werner}}]{The_2025_Roadmap_to_ultrafast_dynamics:_Frontiers_of_theoretical_and_computational_modelling5}%
  \BibitemOpen
  \bibfield  {author} {\bibinfo {author} {\bibfnamefont {F.}~\bibnamefont {Caruso}}, \bibinfo {author} {\bibfnamefont {M.~A.}\ \bibnamefont {Sentef}}, \bibinfo {author} {\bibfnamefont {C.}~\bibnamefont {Attaccalite}}, \bibinfo {author} {\bibfnamefont {M.}~\bibnamefont {Bonitz}}, \bibinfo {author} {\bibfnamefont {C.}~\bibnamefont {Draxl}}, \bibinfo {author} {\bibfnamefont {U.}~\bibnamefont {De~Giovannini}}, \bibinfo {author} {\bibfnamefont {M.}~\bibnamefont {Eckstein}}, \bibinfo {author} {\bibfnamefont {R.}~\bibnamefont {Ernstorfer}}, \bibinfo {author} {\bibfnamefont {M.}~\bibnamefont {Fechner}}, \bibinfo {author} {\bibfnamefont {M.}~\bibnamefont {Grüning}}, \bibinfo {author} {\bibfnamefont {H.}~\bibnamefont {Hübener}}, \bibinfo {author} {\bibfnamefont {J.-P.}\ \bibnamefont {Joost}}, \bibinfo {author} {\bibfnamefont {D.~M.}\ \bibnamefont {Juraschek}}, \bibinfo {author} {\bibfnamefont {C.}~\bibnamefont {Karrasch}}, \bibinfo {author} {\bibfnamefont {D.~M.}\ \bibnamefont {Kennes}}, \bibinfo {author}
  {\bibfnamefont {S.}~\bibnamefont {Latini}}, \bibinfo {author} {\bibfnamefont {I.-T.}\ \bibnamefont {Lu}}, \bibinfo {author} {\bibfnamefont {O.}~\bibnamefont {Neufeld}}, \bibinfo {author} {\bibfnamefont {E.}~\bibnamefont {Perfetto}}, \bibinfo {author} {\bibfnamefont {L.}~\bibnamefont {Rettig}}, \bibinfo {author} {\bibfnamefont {R.~R.}\ \bibnamefont {Pela}}, \bibinfo {author} {\bibfnamefont {A.}~\bibnamefont {Rubio}}, \bibinfo {author} {\bibfnamefont {J.}~\bibnamefont {Rudzinski}}, \bibinfo {author} {\bibfnamefont {M.}~\bibnamefont {Ruggenthaler}}, \bibinfo {author} {\bibfnamefont {D.}~\bibnamefont {Sangalli}}, \bibinfo {author} {\bibfnamefont {M.}~\bibnamefont {Schüler}}, \bibinfo {author} {\bibfnamefont {S.}~\bibnamefont {Shallcross}}, \bibinfo {author} {\bibfnamefont {S.}~\bibnamefont {Sharma}}, \bibinfo {author} {\bibfnamefont {G.}~\bibnamefont {Stefanucci}},\ and\ \bibinfo {author} {\bibfnamefont {P.}~\bibnamefont {Werner}},\ }\bibfield  {title} {\bibinfo {title} {The 2025 roadmap to ultrafast dynamics:
  Frontiers of theoretical and computational modelling},\ }\href {http://iopscience.iop.org/article/10.1088/2515-7639/ae1165} {\bibfield  {journal} {\bibinfo  {journal} {Journal of Physics: Materials}\ } (\bibinfo {year} {2025})}\BibitemShut {NoStop}%
\bibitem [{\citenamefont {Oka}\ and\ \citenamefont {Aoki}(2009)}]{Photovoltaic_Hall_effect_in_graphene_oka_aoki_2009}%
  \BibitemOpen
  \bibfield  {author} {\bibinfo {author} {\bibfnamefont {T.}~\bibnamefont {Oka}}\ and\ \bibinfo {author} {\bibfnamefont {H.}~\bibnamefont {Aoki}},\ }\bibfield  {title} {\bibinfo {title} {Photovoltaic {Hall} effect in graphene},\ }\href {https://doi.org/10.1103/physrevb.79.081406} {\bibfield  {journal} {\bibinfo  {journal} {Physical Review B}\ }\textbf {\bibinfo {volume} {79}},\ \bibinfo {pages} {081406} (\bibinfo {year} {2009})}\BibitemShut {NoStop}%
\bibitem [{\citenamefont {McIver}\ \emph {et~al.}(2019)\citenamefont {McIver}, \citenamefont {Schulte}, \citenamefont {Stein}, \citenamefont {Matsuyama}, \citenamefont {Jotzu}, \citenamefont {Meier},\ and\ \citenamefont {Cavalleri}}]{Light_induced_anomalous_Hall_effect_in_graphene_mciver_schulte_stein_matsuyama_jotzu_meier_cavalleri_2019}%
  \BibitemOpen
  \bibfield  {author} {\bibinfo {author} {\bibfnamefont {J.~W.}\ \bibnamefont {McIver}}, \bibinfo {author} {\bibfnamefont {B.}~\bibnamefont {Schulte}}, \bibinfo {author} {\bibfnamefont {F.-U.}\ \bibnamefont {Stein}}, \bibinfo {author} {\bibfnamefont {T.}~\bibnamefont {Matsuyama}}, \bibinfo {author} {\bibfnamefont {G.}~\bibnamefont {Jotzu}}, \bibinfo {author} {\bibfnamefont {G.}~\bibnamefont {Meier}},\ and\ \bibinfo {author} {\bibfnamefont {A.}~\bibnamefont {Cavalleri}},\ }\bibfield  {title} {\bibinfo {title} {Light-induced anomalous {Hall} effect in graphene},\ }\href {https://doi.org/10.1038/s41567-019-0698-y} {\bibfield  {journal} {\bibinfo  {journal} {Nature Physics}\ }\textbf {\bibinfo {volume} {16}},\ \bibinfo {pages} {38–41} (\bibinfo {year} {2019})}\BibitemShut {NoStop}%
\bibitem [{\citenamefont {Kitagawa}\ \emph {et~al.}(2011)\citenamefont {Kitagawa}, \citenamefont {Oka}, \citenamefont {Brataas}, \citenamefont {Fu},\ and\ \citenamefont {Demler}}]{Transport_properties_of_nonequilibrium_systems_under_the_application_of_light:_Photoinduced_quantum_Hall_insulators_without_Landau_levels}%
  \BibitemOpen
  \bibfield  {author} {\bibinfo {author} {\bibfnamefont {T.}~\bibnamefont {Kitagawa}}, \bibinfo {author} {\bibfnamefont {T.}~\bibnamefont {Oka}}, \bibinfo {author} {\bibfnamefont {A.}~\bibnamefont {Brataas}}, \bibinfo {author} {\bibfnamefont {L.}~\bibnamefont {Fu}},\ and\ \bibinfo {author} {\bibfnamefont {E.}~\bibnamefont {Demler}},\ }\bibfield  {title} {\bibinfo {title} {Transport properties of nonequilibrium systems under the application of light: Photoinduced quantum hall insulators without landau levels},\ }\href {https://doi.org/10.1103/PhysRevB.84.235108} {\bibfield  {journal} {\bibinfo  {journal} {Phys. Rev. B}\ }\textbf {\bibinfo {volume} {84}},\ \bibinfo {pages} {235108} (\bibinfo {year} {2011})}\BibitemShut {NoStop}%
\bibitem [{\citenamefont {Usaj}\ \emph {et~al.}(2014)\citenamefont {Usaj}, \citenamefont {Perez-Piskunow}, \citenamefont {Foa~Torres},\ and\ \citenamefont {Balseiro}}]{Irradiated_graphene_as_a_tunable_Floquet_topological_insulator_usaj_perez-piskunow_foa_torres_balseiro_2014}%
  \BibitemOpen
  \bibfield  {author} {\bibinfo {author} {\bibfnamefont {G.}~\bibnamefont {Usaj}}, \bibinfo {author} {\bibfnamefont {P.~M.}\ \bibnamefont {Perez-Piskunow}}, \bibinfo {author} {\bibfnamefont {L.~E.~F.}\ \bibnamefont {Foa~Torres}},\ and\ \bibinfo {author} {\bibfnamefont {C.~A.}\ \bibnamefont {Balseiro}},\ }\bibfield  {title} {\bibinfo {title} {Irradiated graphene as a tunable {Floquet} topological insulator},\ }\href {https://doi.org/10.1103/physrevb.90.115423} {\bibfield  {journal} {\bibinfo  {journal} {Physical Review B}\ }\textbf {\bibinfo {volume} {90}},\ \bibinfo {pages} {115423} (\bibinfo {year} {2014})}\BibitemShut {NoStop}%
\bibitem [{\citenamefont {Lindner}\ \emph {et~al.}(2011)\citenamefont {Lindner}, \citenamefont {Refael},\ and\ \citenamefont {Galitski}}]{Floquet_topological_insulator_in_semiconductor_quantum_wells_lindner_refael_galitski_2011}%
  \BibitemOpen
  \bibfield  {author} {\bibinfo {author} {\bibfnamefont {N.~H.}\ \bibnamefont {Lindner}}, \bibinfo {author} {\bibfnamefont {G.}~\bibnamefont {Refael}},\ and\ \bibinfo {author} {\bibfnamefont {V.}~\bibnamefont {Galitski}},\ }\bibfield  {title} {\bibinfo {title} {{Floquet} topological insulator in semiconductor quantum wells},\ }\href {https://doi.org/10.1038/nphys1926} {\bibfield  {journal} {\bibinfo  {journal} {Nature Physics}\ }\textbf {\bibinfo {volume} {7}},\ \bibinfo {pages} {490–495} (\bibinfo {year} {2011})}\BibitemShut {NoStop}%
\bibitem [{\citenamefont {Huber}\ \emph {et~al.}(2026)\citenamefont {Huber}, \citenamefont {Kuhlbrodt}, \citenamefont {Anderson}, \citenamefont {Li}, \citenamefont {Watanabe}, \citenamefont {Taniguchi}, \citenamefont {Kroner}, \citenamefont {Xu}, \citenamefont {Imamoğlu},\ and\ \citenamefont {Smoleński}}]{Optical_control_over_topological_Chern_number_in_moire_materials}%
  \BibitemOpen
  \bibfield  {author} {\bibinfo {author} {\bibfnamefont {O.}~\bibnamefont {Huber}}, \bibinfo {author} {\bibfnamefont {K.}~\bibnamefont {Kuhlbrodt}}, \bibinfo {author} {\bibfnamefont {E.}~\bibnamefont {Anderson}}, \bibinfo {author} {\bibfnamefont {W.}~\bibnamefont {Li}}, \bibinfo {author} {\bibfnamefont {K.}~\bibnamefont {Watanabe}}, \bibinfo {author} {\bibfnamefont {T.}~\bibnamefont {Taniguchi}}, \bibinfo {author} {\bibfnamefont {M.}~\bibnamefont {Kroner}}, \bibinfo {author} {\bibfnamefont {X.}~\bibnamefont {Xu}}, \bibinfo {author} {\bibfnamefont {A.}~\bibnamefont {Imamoğlu}},\ and\ \bibinfo {author} {\bibfnamefont {T.}~\bibnamefont {Smoleński}},\ }\bibfield  {title} {\bibinfo {title} {Optical control over topological chern number in moiré materials},\ }\href {https://doi.org/https://doi.org/10.1038/s41586-025-09851-w} {\bibfield  {journal} {\bibinfo  {journal} {Nature}\ }\textbf {\bibinfo {volume} {649}},\ \bibinfo {pages} {1153–1158} (\bibinfo {year} {2026})}\BibitemShut {NoStop}%
\bibitem [{\citenamefont {Genske}\ and\ \citenamefont {Rosch}(2015)}]{Floquet-Boltzmann_equation_for_periodically_driven_Fermi_systems}%
  \BibitemOpen
  \bibfield  {author} {\bibinfo {author} {\bibfnamefont {M.}~\bibnamefont {Genske}}\ and\ \bibinfo {author} {\bibfnamefont {A.}~\bibnamefont {Rosch}},\ }\bibfield  {title} {\bibinfo {title} {Floquet-boltzmann equation for periodically driven fermi systems},\ }\href {https://doi.org/10.1103/PhysRevA.92.062108} {\bibfield  {journal} {\bibinfo  {journal} {Phys. Rev. A}\ }\textbf {\bibinfo {volume} {92}},\ \bibinfo {pages} {062108} (\bibinfo {year} {2015})}\BibitemShut {NoStop}%
\bibitem [{\citenamefont {Seetharam}\ \emph {et~al.}(2015)\citenamefont {Seetharam}, \citenamefont {Bardyn}, \citenamefont {Lindner}, \citenamefont {Rudner},\ and\ \citenamefont {Refael}}]{Controlled_Population_of_Floquet_Bloch_States_via_Coupling_to_Bose_and_Fermi_Baths}%
  \BibitemOpen
  \bibfield  {author} {\bibinfo {author} {\bibfnamefont {K.~I.}\ \bibnamefont {Seetharam}}, \bibinfo {author} {\bibfnamefont {C.-E.}\ \bibnamefont {Bardyn}}, \bibinfo {author} {\bibfnamefont {N.~H.}\ \bibnamefont {Lindner}}, \bibinfo {author} {\bibfnamefont {M.~S.}\ \bibnamefont {Rudner}},\ and\ \bibinfo {author} {\bibfnamefont {G.}~\bibnamefont {Refael}},\ }\bibfield  {title} {\bibinfo {title} {Controlled population of floquet-bloch states via coupling to bose and fermi baths},\ }\href {https://doi.org/10.1103/PhysRevX.5.041050} {\bibfield  {journal} {\bibinfo  {journal} {Phys. Rev. X}\ }\textbf {\bibinfo {volume} {5}},\ \bibinfo {pages} {041050} (\bibinfo {year} {2015})}\BibitemShut {NoStop}%
\bibitem [{\citenamefont {Esin}\ \emph {et~al.}(2018)\citenamefont {Esin}, \citenamefont {Rudner}, \citenamefont {Refael},\ and\ \citenamefont {Lindner}}]{Quantized_transport_and_steady_states_of_Floquet_topological_insulators}%
  \BibitemOpen
  \bibfield  {author} {\bibinfo {author} {\bibfnamefont {I.}~\bibnamefont {Esin}}, \bibinfo {author} {\bibfnamefont {M.~S.}\ \bibnamefont {Rudner}}, \bibinfo {author} {\bibfnamefont {G.}~\bibnamefont {Refael}},\ and\ \bibinfo {author} {\bibfnamefont {N.~H.}\ \bibnamefont {Lindner}},\ }\bibfield  {title} {\bibinfo {title} {Quantized transport and steady states of floquet topological insulators},\ }\href {https://doi.org/10.1103/PhysRevB.97.245401} {\bibfield  {journal} {\bibinfo  {journal} {Phys. Rev. B}\ }\textbf {\bibinfo {volume} {97}},\ \bibinfo {pages} {245401} (\bibinfo {year} {2018})}\BibitemShut {NoStop}%
\bibitem [{\citenamefont {Seetharam}\ \emph {et~al.}(2019)\citenamefont {Seetharam}, \citenamefont {Bardyn}, \citenamefont {Lindner}, \citenamefont {Rudner},\ and\ \citenamefont {Refael}}]{Steady_states_of_interacting_Floquet_insulators}%
  \BibitemOpen
  \bibfield  {author} {\bibinfo {author} {\bibfnamefont {K.~I.}\ \bibnamefont {Seetharam}}, \bibinfo {author} {\bibfnamefont {C.-E.}\ \bibnamefont {Bardyn}}, \bibinfo {author} {\bibfnamefont {N.~H.}\ \bibnamefont {Lindner}}, \bibinfo {author} {\bibfnamefont {M.~S.}\ \bibnamefont {Rudner}},\ and\ \bibinfo {author} {\bibfnamefont {G.}~\bibnamefont {Refael}},\ }\bibfield  {title} {\bibinfo {title} {Steady states of interacting floquet insulators},\ }\href {https://doi.org/10.1103/PhysRevB.99.014307} {\bibfield  {journal} {\bibinfo  {journal} {Phys. Rev. B}\ }\textbf {\bibinfo {volume} {99}},\ \bibinfo {pages} {014307} (\bibinfo {year} {2019})}\BibitemShut {NoStop}%
\bibitem [{\citenamefont {Liu}\ \emph {et~al.}(2025)\citenamefont {Liu}, \citenamefont {Yang}, \citenamefont {Gaertner}, \citenamefont {Huckabee}, \citenamefont {Suslov}, \citenamefont {Refael}, \citenamefont {Nathan}, \citenamefont {Lewandowski}, \citenamefont {Luis}, \citenamefont {Esin}, \citenamefont {Barbara},\ and\ \citenamefont {Kalugin}}]{Signatures_of_Floquet_electronic_steady_states_in_graphene_under_continuous‑wave_mid‑infrared_irradiation_liu_yang_gaertner_huckabee_suslov_refael_nathan_lewandowski_luis_iliya_esin_et_al_2025}%
  \BibitemOpen
  \bibfield  {author} {\bibinfo {author} {\bibfnamefont {Y.}~\bibnamefont {Liu}}, \bibinfo {author} {\bibfnamefont {C.}~\bibnamefont {Yang}}, \bibinfo {author} {\bibfnamefont {G.}~\bibnamefont {Gaertner}}, \bibinfo {author} {\bibfnamefont {J.}~\bibnamefont {Huckabee}}, \bibinfo {author} {\bibfnamefont {A.~V.}\ \bibnamefont {Suslov}}, \bibinfo {author} {\bibfnamefont {G.}~\bibnamefont {Refael}}, \bibinfo {author} {\bibfnamefont {F.}~\bibnamefont {Nathan}}, \bibinfo {author} {\bibfnamefont {C.}~\bibnamefont {Lewandowski}}, \bibinfo {author} {\bibnamefont {Luis}}, \bibinfo {author} {\bibfnamefont {I.}~\bibnamefont {Esin}}, \bibinfo {author} {\bibfnamefont {P.}~\bibnamefont {Barbara}},\ and\ \bibinfo {author} {\bibfnamefont {N.~G.}\ \bibnamefont {Kalugin}},\ }\bibfield  {title} {\bibinfo {title} {Signatures of floquet electronic steady states in graphene under continuous‑wave mid‑infrared irradiation},\ }\href {https://doi.org/10.1038/s41467-025-57335-2} {\bibfield  {journal} {\bibinfo  {journal} {Nature
  Communications}\ }\textbf {\bibinfo {volume} {16}},\ \bibinfo {pages} {57335} (\bibinfo {year} {2025})}\BibitemShut {NoStop}%
\bibitem [{\citenamefont {Perfetto}\ \emph {et~al.}(2022)\citenamefont {Perfetto}, \citenamefont {Pavlyukh},\ and\ \citenamefont {Stefanucci}}]{Real-Time_$GW$:_Toward_an_Ab_Initio_Description_of_the_Ultrafast_Carrier_and_Exciton_Dynamics_in_Two_Dimensional_Materials}%
  \BibitemOpen
  \bibfield  {author} {\bibinfo {author} {\bibfnamefont {E.}~\bibnamefont {Perfetto}}, \bibinfo {author} {\bibfnamefont {Y.}~\bibnamefont {Pavlyukh}},\ and\ \bibinfo {author} {\bibfnamefont {G.}~\bibnamefont {Stefanucci}},\ }\bibfield  {title} {\bibinfo {title} {Real-time $gw$: Toward an ab initio description of the ultrafast carrier and exciton dynamics in two-dimensional materials},\ }\href {https://doi.org/10.1103/PhysRevLett.128.016801} {\bibfield  {journal} {\bibinfo  {journal} {Phys. Rev. Lett.}\ }\textbf {\bibinfo {volume} {128}},\ \bibinfo {pages} {016801} (\bibinfo {year} {2022})}\BibitemShut {NoStop}%
\bibitem [{\citenamefont {Balzer}\ and\ \citenamefont {Bonitz}(2023)}]{Nonequilibrium_Green’s_Functions_Approach_to_Inhomogeneous_Systems}%
  \BibitemOpen
  \bibfield  {author} {\bibinfo {author} {\bibfnamefont {K.}~\bibnamefont {Balzer}}\ and\ \bibinfo {author} {\bibfnamefont {M.}~\bibnamefont {Bonitz}},\ }\href {https://doi.org/https://doi.org/10.1007-978-3-642-35082-5} {\bibinfo {title} {Nonequilibrium green’s functions approach to inhomogeneous systems}} (\bibinfo {year} {2023})\BibitemShut {NoStop}%
\bibitem [{\citenamefont {Schlünzen}\ \emph {et~al.}(2019)\citenamefont {Schlünzen}, \citenamefont {Hermanns}, \citenamefont {Scharnke},\ and\ \citenamefont {Bonitz}}]{Ultrafast_dynamics_of_strongly_correlated_fermions_nonequilibrium_Green_functions_and_selfenergy_approximations}%
  \BibitemOpen
  \bibfield  {author} {\bibinfo {author} {\bibfnamefont {N.}~\bibnamefont {Schlünzen}}, \bibinfo {author} {\bibfnamefont {S.}~\bibnamefont {Hermanns}}, \bibinfo {author} {\bibfnamefont {M.}~\bibnamefont {Scharnke}},\ and\ \bibinfo {author} {\bibfnamefont {M.}~\bibnamefont {Bonitz}},\ }\bibfield  {title} {\bibinfo {title} {Ultrafast dynamics of strongly correlated fermions—nonequilibrium green functions and selfenergy approximations},\ }\href {https://doi.org/https://doi.org/10.1088/1361-648x/ab2d32} {\bibfield  {journal} {\bibinfo  {journal} {Journal of Physics: Condensed Matter}\ }\textbf {\bibinfo {volume} {32}},\ \bibinfo {pages} {103001} (\bibinfo {year} {2019})}\BibitemShut {NoStop}%
\bibitem [{\citenamefont {Marini}\ \emph {et~al.}(2009)\citenamefont {Marini}, \citenamefont {Hogan}, \citenamefont {Gr{\"u}ning},\ and\ \citenamefont {Varsano}}]{Yambo:_an_ab_initio_tool_for_excited_state_calculations}%
  \BibitemOpen
  \bibfield  {author} {\bibinfo {author} {\bibfnamefont {A.}~\bibnamefont {Marini}}, \bibinfo {author} {\bibfnamefont {C.}~\bibnamefont {Hogan}}, \bibinfo {author} {\bibfnamefont {M.}~\bibnamefont {Gr{\"u}ning}},\ and\ \bibinfo {author} {\bibfnamefont {D.}~\bibnamefont {Varsano}},\ }\bibfield  {title} {\bibinfo {title} {Yambo: an ab initio tool for excited state calculations},\ }\href@noop {} {\bibfield  {journal} {\bibinfo  {journal} {Computer Physics Communications}\ }\textbf {\bibinfo {volume} {180}},\ \bibinfo {pages} {1392} (\bibinfo {year} {2009})}\BibitemShut {NoStop}%
\bibitem [{\citenamefont {Sch\"uler}\ \emph {et~al.}(2019)\citenamefont {Sch\"uler}, \citenamefont {Budich},\ and\ \citenamefont {Werner}}]{Quench_dynamics_and_Hall_response_of_interacting_Chern_insulators}%
  \BibitemOpen
  \bibfield  {author} {\bibinfo {author} {\bibfnamefont {M.}~\bibnamefont {Sch\"uler}}, \bibinfo {author} {\bibfnamefont {J.~C.}\ \bibnamefont {Budich}},\ and\ \bibinfo {author} {\bibfnamefont {P.}~\bibnamefont {Werner}},\ }\bibfield  {title} {\bibinfo {title} {Quench dynamics and hall response of interacting chern insulators},\ }\href {https://doi.org/10.1103/PhysRevB.100.041101} {\bibfield  {journal} {\bibinfo  {journal} {Phys. Rev. B}\ }\textbf {\bibinfo {volume} {100}},\ \bibinfo {pages} {041101} (\bibinfo {year} {2019})}\BibitemShut {NoStop}%
\bibitem [{\citenamefont {Tancogne-Dejean}\ \emph {et~al.}(2020)\citenamefont {Tancogne-Dejean}, \citenamefont {Oliveira}, \citenamefont {Andrade}, \citenamefont {Appel}, \citenamefont {Borca}, \citenamefont {Le~Breton}, \citenamefont {Buchholz}, \citenamefont {Castro}, \citenamefont {Corni}, \citenamefont {Correa}, \citenamefont {De~Giovannini}, \citenamefont {Delgado}, \citenamefont {Eich}, \citenamefont {Flick}, \citenamefont {Gil}, \citenamefont {Gomez}, \citenamefont {Helbig}, \citenamefont {Hübener}, \citenamefont {Jestädt},\ and\ \citenamefont {Jornet-Somoza}}]{Octopus_a_computational_framework_for_exploring_light-driven_phenomena_and_quantum_dynamics_in_extended_and_finite_systems_tancogne-dejean_oliveira_andrade_appel_borca_le_breton_buchholz_castro_corni_correa_et_al_2020}%
  \BibitemOpen
  \bibfield  {author} {\bibinfo {author} {\bibfnamefont {N.}~\bibnamefont {Tancogne-Dejean}}, \bibinfo {author} {\bibfnamefont {M.~J.~T.}\ \bibnamefont {Oliveira}}, \bibinfo {author} {\bibfnamefont {X.}~\bibnamefont {Andrade}}, \bibinfo {author} {\bibfnamefont {H.}~\bibnamefont {Appel}}, \bibinfo {author} {\bibfnamefont {C.~H.}\ \bibnamefont {Borca}}, \bibinfo {author} {\bibfnamefont {G.}~\bibnamefont {Le~Breton}}, \bibinfo {author} {\bibfnamefont {F.}~\bibnamefont {Buchholz}}, \bibinfo {author} {\bibfnamefont {A.}~\bibnamefont {Castro}}, \bibinfo {author} {\bibfnamefont {S.}~\bibnamefont {Corni}}, \bibinfo {author} {\bibfnamefont {A.~A.}\ \bibnamefont {Correa}}, \bibinfo {author} {\bibfnamefont {U.}~\bibnamefont {De~Giovannini}}, \bibinfo {author} {\bibfnamefont {A.}~\bibnamefont {Delgado}}, \bibinfo {author} {\bibfnamefont {F.~G.}\ \bibnamefont {Eich}}, \bibinfo {author} {\bibfnamefont {J.}~\bibnamefont {Flick}}, \bibinfo {author} {\bibfnamefont {G.}~\bibnamefont {Gil}}, \bibinfo {author} {\bibfnamefont
  {A.}~\bibnamefont {Gomez}}, \bibinfo {author} {\bibfnamefont {N.}~\bibnamefont {Helbig}}, \bibinfo {author} {\bibfnamefont {H.}~\bibnamefont {Hübener}}, \bibinfo {author} {\bibfnamefont {R.}~\bibnamefont {Jestädt}},\ and\ \bibinfo {author} {\bibfnamefont {J.}~\bibnamefont {Jornet-Somoza}},\ }\bibfield  {title} {\bibinfo {title} {Octopus, a computational framework for exploring light-driven phenomena and quantum dynamics in extended and finite systems},\ }\href {https://doi.org/10.1063/1.5142502} {\bibfield  {journal} {\bibinfo  {journal} {The Journal of Chemical Physics}\ }\textbf {\bibinfo {volume} {152}},\ \bibinfo {pages} {124119} (\bibinfo {year} {2020})}\BibitemShut {NoStop}%
\bibitem [{\citenamefont {Choi}\ \emph {et~al.}(2025)\citenamefont {Choi}, \citenamefont {Mogi}, \citenamefont {De~Giovannini}, \citenamefont {Azoury}, \citenamefont {Lv}, \citenamefont {Su}, \citenamefont {Hübener}, \citenamefont {Rubio},\ and\ \citenamefont {Gedik}}]{Observation_of_Floquet_Bloch_states_in_monolayer_graphene_choi_mogi_de_giovannini_azoury_lv_su_hübener_rubio_gedik_2025}%
  \BibitemOpen
  \bibfield  {author} {\bibinfo {author} {\bibfnamefont {D.}~\bibnamefont {Choi}}, \bibinfo {author} {\bibfnamefont {M.}~\bibnamefont {Mogi}}, \bibinfo {author} {\bibfnamefont {U.}~\bibnamefont {De~Giovannini}}, \bibinfo {author} {\bibfnamefont {D.}~\bibnamefont {Azoury}}, \bibinfo {author} {\bibfnamefont {B.}~\bibnamefont {Lv}}, \bibinfo {author} {\bibfnamefont {Y.}~\bibnamefont {Su}}, \bibinfo {author} {\bibfnamefont {H.}~\bibnamefont {Hübener}}, \bibinfo {author} {\bibfnamefont {A.}~\bibnamefont {Rubio}},\ and\ \bibinfo {author} {\bibfnamefont {N.}~\bibnamefont {Gedik}},\ }\bibfield  {title} {\bibinfo {title} {Observation of {Floquet}–{Bloch} states in monolayer graphene},\ }\href {https://doi.org/10.1038/s41567-025-02888-8} {\bibfield  {journal} {\bibinfo  {journal} {Nature Physics}\ ,\ \bibinfo {pages} {02888}} (\bibinfo {year} {2025})}\BibitemShut {NoStop}%
\bibitem [{\citenamefont {Attaccalite}\ \emph {et~al.}(2011)\citenamefont {Attaccalite}, \citenamefont {Gr\"uning},\ and\ \citenamefont {Marini}}]{Real-time_approach_to_the_optical_properties_of_solids_and_nanostructures:_Time_dependent_Bethe_Salpeter_equation}%
  \BibitemOpen
  \bibfield  {author} {\bibinfo {author} {\bibfnamefont {C.}~\bibnamefont {Attaccalite}}, \bibinfo {author} {\bibfnamefont {M.}~\bibnamefont {Gr\"uning}},\ and\ \bibinfo {author} {\bibfnamefont {A.}~\bibnamefont {Marini}},\ }\bibfield  {title} {\bibinfo {title} {Real-time approach to the optical properties of solids and nanostructures: Time-dependent bethe-salpeter equation},\ }\href {https://doi.org/10.1103/PhysRevB.84.245110} {\bibfield  {journal} {\bibinfo  {journal} {Phys. Rev. B}\ }\textbf {\bibinfo {volume} {84}},\ \bibinfo {pages} {245110} (\bibinfo {year} {2011})}\BibitemShut {NoStop}%
\bibitem [{\citenamefont {Ridolfi}\ \emph {et~al.}(2020)\citenamefont {Ridolfi}, \citenamefont {Trevisanutto},\ and\ \citenamefont {Pereira}}]{Expeditious_computation_of_nonlinear_optical_properties_of_arbitrary_order_with_native_electronic_interactions_in_the_time_domain}%
  \BibitemOpen
  \bibfield  {author} {\bibinfo {author} {\bibfnamefont {E.}~\bibnamefont {Ridolfi}}, \bibinfo {author} {\bibfnamefont {P.~E.}\ \bibnamefont {Trevisanutto}},\ and\ \bibinfo {author} {\bibfnamefont {V.~M.}\ \bibnamefont {Pereira}},\ }\bibfield  {title} {\bibinfo {title} {Expeditious computation of nonlinear optical properties of arbitrary order with native electronic interactions in the time domain},\ }\href {https://doi.org/10.1103/PhysRevB.102.245110} {\bibfield  {journal} {\bibinfo  {journal} {Phys. Rev. B}\ }\textbf {\bibinfo {volume} {102}},\ \bibinfo {pages} {245110} (\bibinfo {year} {2020})}\BibitemShut {NoStop}%
\bibitem [{\citenamefont {Chan}\ \emph {et~al.}(2023)\citenamefont {Chan}, \citenamefont {Qiu}, \citenamefont {da~Jornada},\ and\ \citenamefont {Louie}}]{Giant_self-driven_exciton-Floquet_signatures_in_time-resolved_photoemission_spectroscopy_of_MoS<sub>2</sub>_from_time-dependent_GW_approach}%
  \BibitemOpen
  \bibfield  {author} {\bibinfo {author} {\bibfnamefont {Y.-H.}\ \bibnamefont {Chan}}, \bibinfo {author} {\bibfnamefont {D.~Y.}\ \bibnamefont {Qiu}}, \bibinfo {author} {\bibfnamefont {F.~H.}\ \bibnamefont {da~Jornada}},\ and\ \bibinfo {author} {\bibfnamefont {S.~G.}\ \bibnamefont {Louie}},\ }\bibfield  {title} {\bibinfo {title} {Giant self-driven exciton-floquet signatures in time-resolved photoemission spectroscopy of mos<sub>2</sub> from time-dependent gw approach},\ }\href {https://doi.org/10.1073/pnas.2301957120} {\bibfield  {journal} {\bibinfo  {journal} {Proceedings of the National Academy of Sciences}\ }\textbf {\bibinfo {volume} {120}},\ \bibinfo {pages} {e2301957120} (\bibinfo {year} {2023})},\ \Eprint {https://arxiv.org/abs/https://www.pnas.org/doi/pdf/10.1073/pnas.2301957120} {https://www.pnas.org/doi/pdf/10.1073/pnas.2301957120} \BibitemShut {NoStop}%
\bibitem [{\citenamefont {Tanaka}\ and\ \citenamefont {Mochizuki}(2021)}]{Real-time_dynamics_of_the_photoinduced_topological_state_in_the_organic_conductor}%
  \BibitemOpen
  \bibfield  {author} {\bibinfo {author} {\bibfnamefont {Y.}~\bibnamefont {Tanaka}}\ and\ \bibinfo {author} {\bibfnamefont {M.}~\bibnamefont {Mochizuki}},\ }\bibfield  {title} {\bibinfo {title} {Real-time dynamics of the photoinduced topological state in the organic conductor $\ensuremath{\alpha}\text{\ensuremath{-}}{(\mathrm{BEDT}\text{\ensuremath{-}}\mathrm{TTF})}_{2}{\mathrm{i}}_{3}$ under continuous-wave and pulse excitations},\ }\href {https://doi.org/10.1103/PhysRevB.104.085123} {\bibfield  {journal} {\bibinfo  {journal} {Phys. Rev. B}\ }\textbf {\bibinfo {volume} {104}},\ \bibinfo {pages} {085123} (\bibinfo {year} {2021})}\BibitemShut {NoStop}%
\bibitem [{\citenamefont {Veiga}\ \emph {et~al.}(2024)\citenamefont {Veiga}, \citenamefont {João}, \citenamefont {Pinho}, \citenamefont {Pires},\ and\ \citenamefont {Lopes}}]{Unambiguous_simulation_of_diffusive_charge_transport_in_disordered_nanoribbons}%
  \BibitemOpen
  \bibfield  {author} {\bibinfo {author} {\bibfnamefont {H.~P.}\ \bibnamefont {Veiga}}, \bibinfo {author} {\bibfnamefont {S.~M.}\ \bibnamefont {João}}, \bibinfo {author} {\bibfnamefont {J.~M.~A.}\ \bibnamefont {Pinho}}, \bibinfo {author} {\bibfnamefont {J.~P.~S.}\ \bibnamefont {Pires}},\ and\ \bibinfo {author} {\bibfnamefont {J.~M. V.~P.}\ \bibnamefont {Lopes}},\ }\bibfield  {title} {\bibinfo {title} {{Unambiguous simulation of diffusive charge transport in disordered nanoribbons}},\ }\href {https://doi.org/10.21468/SciPostPhys.17.6.149} {\bibfield  {journal} {\bibinfo  {journal} {SciPost Phys.}\ }\textbf {\bibinfo {volume} {17}},\ \bibinfo {pages} {149} (\bibinfo {year} {2024})}\BibitemShut {NoStop}%
\bibitem [{\citenamefont {Pinho}\ \emph {et~al.}(2023)\citenamefont {Pinho}, \citenamefont {Pires}, \citenamefont {Jo\~ao}, \citenamefont {Amorim},\ and\ \citenamefont {Lopes}}]{From_Bloch_Oscillations_to_a_Steady_State_Current_in_Strongly_Biased_Mesoscopic_Devices}%
  \BibitemOpen
  \bibfield  {author} {\bibinfo {author} {\bibfnamefont {J.~M.~A.}\ \bibnamefont {Pinho}}, \bibinfo {author} {\bibfnamefont {J.~P.~S.}\ \bibnamefont {Pires}}, \bibinfo {author} {\bibfnamefont {S.~M.}\ \bibnamefont {Jo\~ao}}, \bibinfo {author} {\bibfnamefont {B.}~\bibnamefont {Amorim}},\ and\ \bibinfo {author} {\bibfnamefont {J.~M. V.~P.}\ \bibnamefont {Lopes}},\ }\bibfield  {title} {\bibinfo {title} {From bloch oscillations to a steady-state current in strongly biased mesoscopic devices},\ }\href {https://doi.org/10.1103/PhysRevB.108.075402} {\bibfield  {journal} {\bibinfo  {journal} {Phys. Rev. B}\ }\textbf {\bibinfo {volume} {108}},\ \bibinfo {pages} {075402} (\bibinfo {year} {2023})}\BibitemShut {NoStop}%
\bibitem [{\citenamefont {Santos~Pires}\ \emph {et~al.}(2020)\citenamefont {Santos~Pires}, \citenamefont {Amorim},\ and\ \citenamefont {Viana Parente~Lopes}}]{Landauer_transport_as_a_quasisteady_state_on_finite_chains_under_unitary_quantum_dynamics}%
  \BibitemOpen
  \bibfield  {author} {\bibinfo {author} {\bibfnamefont {J.~P.}\ \bibnamefont {Santos~Pires}}, \bibinfo {author} {\bibfnamefont {B.}~\bibnamefont {Amorim}},\ and\ \bibinfo {author} {\bibfnamefont {J.~M.}\ \bibnamefont {Viana Parente~Lopes}},\ }\bibfield  {title} {\bibinfo {title} {Landauer transport as a quasisteady state on finite chains under unitary quantum dynamics},\ }\href {https://doi.org/10.1103/PhysRevB.101.104203} {\bibfield  {journal} {\bibinfo  {journal} {Phys. Rev. B}\ }\textbf {\bibinfo {volume} {101}},\ \bibinfo {pages} {104203} (\bibinfo {year} {2020})}\BibitemShut {NoStop}%
\bibitem [{\citenamefont {Veiga}\ \emph {et~al.}(2025)\citenamefont {Veiga}, \citenamefont {Pinheiro}, \citenamefont {Pires},\ and\ \citenamefont {Lopes}}]{Markov_Inequality_as_a_Tool_for_Linear_Scaling_Estimation_of_Local_Observables}%
  \BibitemOpen
  \bibfield  {author} {\bibinfo {author} {\bibfnamefont {H.}~\bibnamefont {Veiga}}, \bibinfo {author} {\bibfnamefont {D.}~\bibnamefont {Pinheiro}}, \bibinfo {author} {\bibfnamefont {J.}~\bibnamefont {Pires}},\ and\ \bibinfo {author} {\bibfnamefont {J.}~\bibnamefont {Lopes}},\ }\bibfield  {title} {\bibinfo {title} {Markov inequality as a tool for linear-scaling estimation of local observables},\ }\href@noop {} {\bibfield  {journal} {\bibinfo  {journal} {arXiv preprint arXiv:2510.21688}\ } (\bibinfo {year} {2025})}\BibitemShut {NoStop}%
\bibitem [{\citenamefont {Shao}\ \emph {et~al.}(2021)\citenamefont {Shao}, \citenamefont {Sacramento},\ and\ \citenamefont {Mondaini}}]{Photoinduced_anomalous_Hall_effect_in_the_interacting_Haldane_model:_Targeting_topological_states_with_pump_pulses}%
  \BibitemOpen
  \bibfield  {author} {\bibinfo {author} {\bibfnamefont {C.}~\bibnamefont {Shao}}, \bibinfo {author} {\bibfnamefont {P.~D.}\ \bibnamefont {Sacramento}},\ and\ \bibinfo {author} {\bibfnamefont {R.}~\bibnamefont {Mondaini}},\ }\bibfield  {title} {\bibinfo {title} {Photoinduced anomalous hall effect in the interacting haldane model: Targeting topological states with pump pulses},\ }\href {https://doi.org/10.1103/PhysRevB.104.125129} {\bibfield  {journal} {\bibinfo  {journal} {Phys. Rev. B}\ }\textbf {\bibinfo {volume} {104}},\ \bibinfo {pages} {125129} (\bibinfo {year} {2021})}\BibitemShut {NoStop}%
\bibitem [{\citenamefont {Fan}\ \emph {et~al.}(2020)\citenamefont {Fan}, \citenamefont {Garcia}, \citenamefont {Cummings}, \citenamefont {Barrios-Vargas}, \citenamefont {Panhans}, \citenamefont {Harju}, \citenamefont {Ortmann},\ and\ \citenamefont {Roche}}]{Linear_scaling_quantum_transport_methodologies_fan_garcia_cummings_barrios-vargas_panhans_harju_ortmann_roche_2020}%
  \BibitemOpen
  \bibfield  {author} {\bibinfo {author} {\bibfnamefont {Z.}~\bibnamefont {Fan}}, \bibinfo {author} {\bibfnamefont {J.~H.}\ \bibnamefont {Garcia}}, \bibinfo {author} {\bibfnamefont {A.~W.}\ \bibnamefont {Cummings}}, \bibinfo {author} {\bibfnamefont {J.~E.}\ \bibnamefont {Barrios-Vargas}}, \bibinfo {author} {\bibfnamefont {M.}~\bibnamefont {Panhans}}, \bibinfo {author} {\bibfnamefont {A.}~\bibnamefont {Harju}}, \bibinfo {author} {\bibfnamefont {F.}~\bibnamefont {Ortmann}},\ and\ \bibinfo {author} {\bibfnamefont {S.}~\bibnamefont {Roche}},\ }\bibfield  {title} {\bibinfo {title} {Linear scaling quantum transport methodologies},\ }\href {https://doi.org/10.1016/j.physrep.2020.12.001} {\bibfield  {journal} {\bibinfo  {journal} {Physics Reports}\ }\textbf {\bibinfo {volume} {903}},\ \bibinfo {pages} {1–69} (\bibinfo {year} {2020})}\BibitemShut {NoStop}%
\bibitem [{\citenamefont {Jo{\~{a}}o}\ \emph {et~al.}(2020)\citenamefont {Jo{\~{a}}o}, \citenamefont {An{\dj}elkovi{\'{c}}}, \citenamefont {Covaci}, \citenamefont {Rappoport}, \citenamefont {Lopes},\ and\ \citenamefont {Ferreira}}]{KITE}%
  \BibitemOpen
  \bibfield  {author} {\bibinfo {author} {\bibfnamefont {S.~M.}\ \bibnamefont {Jo{\~{a}}o}}, \bibinfo {author} {\bibfnamefont {M.}~\bibnamefont {An{\dj}elkovi{\'{c}}}}, \bibinfo {author} {\bibfnamefont {L.}~\bibnamefont {Covaci}}, \bibinfo {author} {\bibfnamefont {T.~G.}\ \bibnamefont {Rappoport}}, \bibinfo {author} {\bibfnamefont {J.~M. V.~P.}\ \bibnamefont {Lopes}},\ and\ \bibinfo {author} {\bibfnamefont {A.}~\bibnamefont {Ferreira}},\ }\bibfield  {title} {\bibinfo {title} {{KITE}: high-performance accurate modelling of electronic structure and response functions of large molecules, disordered crystals and heterostructures},\ }\href {https://doi.org/10.1098/rsos.191809} {\bibfield  {journal} {\bibinfo  {journal} {Royal Society Open Science}\ }\textbf {\bibinfo {volume} {7}},\ \bibinfo {pages} {191809} (\bibinfo {year} {2020})}\BibitemShut {NoStop}%
\bibitem [{\citenamefont {Guerrero}\ \emph {et~al.}(2025)\citenamefont {Guerrero}, \citenamefont {Nguyen}, \citenamefont {Romeral}, \citenamefont {Cummings}, \citenamefont {Garcia}, \citenamefont {Charlier},\ and\ \citenamefont {Roche}}]{Disorder_Induced_Delocalization_in_Magic_Angle_Twisted_Bilayer_Graphene_guerrero_nguyen_romeral_cummings_garcia_charlier_roche_2025}%
  \BibitemOpen
  \bibfield  {author} {\bibinfo {author} {\bibfnamefont {P.~A.}\ \bibnamefont {Guerrero}}, \bibinfo {author} {\bibfnamefont {V.-H.}\ \bibnamefont {Nguyen}}, \bibinfo {author} {\bibfnamefont {J.~M.}\ \bibnamefont {Romeral}}, \bibinfo {author} {\bibfnamefont {A.~W.}\ \bibnamefont {Cummings}}, \bibinfo {author} {\bibfnamefont {J.-H.}\ \bibnamefont {Garcia}}, \bibinfo {author} {\bibfnamefont {J.-C.}\ \bibnamefont {Charlier}},\ and\ \bibinfo {author} {\bibfnamefont {S.}~\bibnamefont {Roche}},\ }\bibfield  {title} {\bibinfo {title} {Disorder-induced delocalization in magic-angle twisted bilayer graphene},\ }\href {https://doi.org/10.1103/physrevlett.134.126301} {\bibfield  {journal} {\bibinfo  {journal} {Physical Review Letters}\ }\textbf {\bibinfo {volume} {134}},\ \bibinfo {pages} {126301} (\bibinfo {year} {2025})}\BibitemShut {NoStop}%
\bibitem [{\citenamefont {Romeral}\ \emph {et~al.}(2025)\citenamefont {Romeral}, \citenamefont {Cummings},\ and\ \citenamefont {Roche}}]{Scaling_of_the_integrated_quantum_metric_in_disordered_topological_phases_romeral_cummings_roche_2025}%
  \BibitemOpen
  \bibfield  {author} {\bibinfo {author} {\bibfnamefont {J.~M.}\ \bibnamefont {Romeral}}, \bibinfo {author} {\bibfnamefont {A.~W.}\ \bibnamefont {Cummings}},\ and\ \bibinfo {author} {\bibfnamefont {S.}~\bibnamefont {Roche}},\ }\bibfield  {title} {\bibinfo {title} {Scaling of the integrated quantum metric in disordered topological phases},\ }\href {https://doi.org/10.1103/physrevb.111.134201} {\bibfield  {journal} {\bibinfo  {journal} {Physical Review B}\ }\textbf {\bibinfo {volume} {111}},\ \bibinfo {pages} {134201} (\bibinfo {year} {2025})}\BibitemShut {NoStop}%
\bibitem [{\citenamefont {Galvani}\ \emph {et~al.}(2024)\citenamefont {Galvani}, \citenamefont {Hamze}, \citenamefont {Caputo}, \citenamefont {Kaya}, \citenamefont {Dubois}, \citenamefont {Colombo}, \citenamefont {Nguyen}, \citenamefont {Shin}, \citenamefont {Shin}, \citenamefont {Charlier},\ and\ \citenamefont {Roche}}]{Exploring_dielectric_properties_in_atomistic_models_of_amorphous_boron_nitride_galvani_hamze_caputo_kaya_dubois_colombo_nguyen_shin_shin_charlier_et_al_2024}%
  \BibitemOpen
  \bibfield  {author} {\bibinfo {author} {\bibfnamefont {T.}~\bibnamefont {Galvani}}, \bibinfo {author} {\bibfnamefont {A.~K.}\ \bibnamefont {Hamze}}, \bibinfo {author} {\bibfnamefont {L.}~\bibnamefont {Caputo}}, \bibinfo {author} {\bibfnamefont {O.}~\bibnamefont {Kaya}}, \bibinfo {author} {\bibfnamefont {S.~M.-M.}\ \bibnamefont {Dubois}}, \bibinfo {author} {\bibfnamefont {L.}~\bibnamefont {Colombo}}, \bibinfo {author} {\bibfnamefont {V.-H.}\ \bibnamefont {Nguyen}}, \bibinfo {author} {\bibfnamefont {Y.}~\bibnamefont {Shin}}, \bibinfo {author} {\bibfnamefont {H.-J.}\ \bibnamefont {Shin}}, \bibinfo {author} {\bibfnamefont {J.-C.}\ \bibnamefont {Charlier}},\ and\ \bibinfo {author} {\bibfnamefont {S.}~\bibnamefont {Roche}},\ }\bibfield  {title} {\bibinfo {title} {Exploring dielectric properties in atomistic models of amorphous boron nitride},\ }\href {https://doi.org/10.1088/2515-7639/ad4c06} {\bibfield  {journal} {\bibinfo  {journal} {Journal of Physics Materials}\ }\textbf {\bibinfo {volume} {7}},\ \bibinfo
  {pages} {035003–035003} (\bibinfo {year} {2024})}\BibitemShut {NoStop}%
\bibitem [{\citenamefont {Canonico}\ \emph {et~al.}(2026)\citenamefont {Canonico}, \citenamefont {Roche},\ and\ \citenamefont {Cummings}}]{Real_Time_Out_of_Equilibrium_Quantum_Dynamics_in_Disordered_Materials_canonico_roche_cummings_2024}%
  \BibitemOpen
  \bibfield  {author} {\bibinfo {author} {\bibfnamefont {L.~M.}\ \bibnamefont {Canonico}}, \bibinfo {author} {\bibfnamefont {S.}~\bibnamefont {Roche}},\ and\ \bibinfo {author} {\bibfnamefont {A.~W.}\ \bibnamefont {Cummings}},\ }\bibfield  {title} {\bibinfo {title} {Real-time out-of-equilibrium quantum dynamics in disordered materials},\ }\href {https://doi.org/10.1103/28hd-pwhv} {\bibfield  {journal} {\bibinfo  {journal} {Phys. Rev. Lett.}\ }\textbf {\bibinfo {volume} {136}},\ \bibinfo {pages} {016903} (\bibinfo {year} {2026})}\BibitemShut {NoStop}%
\bibitem [{\citenamefont {Sato}\ \emph {et~al.}(2019)\citenamefont {Sato}, \citenamefont {McIver}, \citenamefont {Nuske}, \citenamefont {Tang}, \citenamefont {Jotzu}, \citenamefont {Schulte}, \citenamefont {H\"ubener}, \citenamefont {De~Giovannini}, \citenamefont {Mathey}, \citenamefont {Sentef}, \citenamefont {Cavalleri},\ and\ \citenamefont {Rubio}}]{Microscopic_theory_for_the_light_induced_anomalous_Hall_effect_in_graphene}%
  \BibitemOpen
  \bibfield  {author} {\bibinfo {author} {\bibfnamefont {S.~A.}\ \bibnamefont {Sato}}, \bibinfo {author} {\bibfnamefont {J.~W.}\ \bibnamefont {McIver}}, \bibinfo {author} {\bibfnamefont {M.}~\bibnamefont {Nuske}}, \bibinfo {author} {\bibfnamefont {P.}~\bibnamefont {Tang}}, \bibinfo {author} {\bibfnamefont {G.}~\bibnamefont {Jotzu}}, \bibinfo {author} {\bibfnamefont {B.}~\bibnamefont {Schulte}}, \bibinfo {author} {\bibfnamefont {H.}~\bibnamefont {H\"ubener}}, \bibinfo {author} {\bibfnamefont {U.}~\bibnamefont {De~Giovannini}}, \bibinfo {author} {\bibfnamefont {L.}~\bibnamefont {Mathey}}, \bibinfo {author} {\bibfnamefont {M.~A.}\ \bibnamefont {Sentef}}, \bibinfo {author} {\bibfnamefont {A.}~\bibnamefont {Cavalleri}},\ and\ \bibinfo {author} {\bibfnamefont {A.}~\bibnamefont {Rubio}},\ }\bibfield  {title} {\bibinfo {title} {Microscopic theory for the light-induced anomalous hall effect in graphene},\ }\href {https://doi.org/10.1103/PhysRevB.99.214302} {\bibfield  {journal} {\bibinfo  {journal} {Phys. Rev. B}\
  }\textbf {\bibinfo {volume} {99}},\ \bibinfo {pages} {214302} (\bibinfo {year} {2019})}\BibitemShut {NoStop}%
\bibitem [{\citenamefont {Merboldt}\ \emph {et~al.}(2025{\natexlab{a}})\citenamefont {Merboldt}, \citenamefont {Schüler}, \citenamefont {Schmitt}, \citenamefont {Bange}, \citenamefont {Bennecke}, \citenamefont {Gadge}, \citenamefont {Pierz}, \citenamefont {Schumacher}, \citenamefont {Momeni}, \citenamefont {Steil}, \citenamefont {Manmana}, \citenamefont {Sentef}, \citenamefont {Reutzel},\ and\ \citenamefont {Mathias}}]{Observation_of_Floquet_states_in_graphene_merboldt_et_al_2025}%
  \BibitemOpen
  \bibfield  {author} {\bibinfo {author} {\bibfnamefont {M.}~\bibnamefont {Merboldt}}, \bibinfo {author} {\bibfnamefont {M.}~\bibnamefont {Schüler}}, \bibinfo {author} {\bibfnamefont {D.}~\bibnamefont {Schmitt}}, \bibinfo {author} {\bibfnamefont {J.~P.}\ \bibnamefont {Bange}}, \bibinfo {author} {\bibfnamefont {W.}~\bibnamefont {Bennecke}}, \bibinfo {author} {\bibfnamefont {K.}~\bibnamefont {Gadge}}, \bibinfo {author} {\bibfnamefont {K.}~\bibnamefont {Pierz}}, \bibinfo {author} {\bibfnamefont {H.~W.}\ \bibnamefont {Schumacher}}, \bibinfo {author} {\bibfnamefont {D.}~\bibnamefont {Momeni}}, \bibinfo {author} {\bibfnamefont {D.}~\bibnamefont {Steil}}, \bibinfo {author} {\bibfnamefont {S.~R.}\ \bibnamefont {Manmana}}, \bibinfo {author} {\bibfnamefont {M.~A.}\ \bibnamefont {Sentef}}, \bibinfo {author} {\bibfnamefont {M.}~\bibnamefont {Reutzel}},\ and\ \bibinfo {author} {\bibfnamefont {S.}~\bibnamefont {Mathias}},\ }\bibfield  {title} {\bibinfo {title} {Observation of {Floquet} states in graphene},\ }\href
  {https://doi.org/10.1038/s41567-025-02889-7} {\bibfield  {journal} {\bibinfo  {journal} {Nature Physics}\ }\textbf {\bibinfo {volume} {21}},\ \bibinfo {pages} {1745} (\bibinfo {year} {2025}{\natexlab{a}})}\BibitemShut {NoStop}%
\bibitem [{\citenamefont {Merboldt}\ \emph {et~al.}(2025{\natexlab{b}})\citenamefont {Merboldt}, \citenamefont {Schüler}, \citenamefont {Schmitt}, \citenamefont {Bange}, \citenamefont {Bennecke}, \citenamefont {Gadge}, \citenamefont {Pierz}, \citenamefont {Schumacher}, \citenamefont {Momeni}, \citenamefont {Steil}, \citenamefont {Manmana}, \citenamefont {Sentef}, \citenamefont {Reutzel},\ and\ \citenamefont {Mathias}}]{Observation_of_Floquet_states_in_graphene_merboldt_schüler_schmitt_bange_bennecke_karun_gadge_pierz_schumacher_momeni_steil_et_al_2025}%
  \BibitemOpen
  \bibfield  {author} {\bibinfo {author} {\bibfnamefont {M.}~\bibnamefont {Merboldt}}, \bibinfo {author} {\bibfnamefont {M.}~\bibnamefont {Schüler}}, \bibinfo {author} {\bibfnamefont {D.}~\bibnamefont {Schmitt}}, \bibinfo {author} {\bibfnamefont {J.~P.}\ \bibnamefont {Bange}}, \bibinfo {author} {\bibfnamefont {W.}~\bibnamefont {Bennecke}}, \bibinfo {author} {\bibfnamefont {K.}~\bibnamefont {Gadge}}, \bibinfo {author} {\bibfnamefont {K.}~\bibnamefont {Pierz}}, \bibinfo {author} {\bibfnamefont {H.~W.}\ \bibnamefont {Schumacher}}, \bibinfo {author} {\bibfnamefont {D.}~\bibnamefont {Momeni}}, \bibinfo {author} {\bibfnamefont {D.}~\bibnamefont {Steil}}, \bibinfo {author} {\bibfnamefont {S.~R.}\ \bibnamefont {Manmana}}, \bibinfo {author} {\bibfnamefont {M.~A.}\ \bibnamefont {Sentef}}, \bibinfo {author} {\bibfnamefont {M.}~\bibnamefont {Reutzel}},\ and\ \bibinfo {author} {\bibfnamefont {S.}~\bibnamefont {Mathias}},\ }\bibfield  {title} {\bibinfo {title} {Observation of floquet states in graphene},\ }\href
  {https://doi.org/10.1038/s41567-025-02889-7} {\bibfield  {journal} {\bibinfo  {journal} {Nature Physics}\ }\textbf {\bibinfo {volume} {21}},\ \bibinfo {pages} {1093} (\bibinfo {year} {2025}{\natexlab{b}})}\BibitemShut {NoStop}%
\bibitem [{\citenamefont {Haldane}(1988)}]{HaldaneModel}%
  \BibitemOpen
  \bibfield  {author} {\bibinfo {author} {\bibfnamefont {F.~D.~M.}\ \bibnamefont {Haldane}},\ }\bibfield  {title} {\bibinfo {title} {Model for a quantum {Hall} effect without landau levels: Condensed-matter realization of the ``parity anomaly''},\ }\href {https://doi.org/10.1103/PhysRevLett.61.2015} {\bibfield  {journal} {\bibinfo  {journal} {Physical Review Letters}\ }\textbf {\bibinfo {volume} {61}},\ \bibinfo {pages} {2015} (\bibinfo {year} {1988})}\BibitemShut {NoStop}%
\bibitem [{\citenamefont {Zhang}\ \emph {et~al.}(2009)\citenamefont {Zhang}, \citenamefont {Tang}, \citenamefont {Girit}, \citenamefont {Hao}, \citenamefont {Martin}, \citenamefont {Zettl}, \citenamefont {Crommie}, \citenamefont {Shen},\ and\ \citenamefont {Wang}}]{Direct_observation_of_a_widely_tunable_bandgap_in_bilayer_graphene_zhang_tang_girit_hao_martin_zettl_crommie_shen_wang_2009}%
  \BibitemOpen
  \bibfield  {author} {\bibinfo {author} {\bibfnamefont {Y.}~\bibnamefont {Zhang}}, \bibinfo {author} {\bibfnamefont {T.-T.}\ \bibnamefont {Tang}}, \bibinfo {author} {\bibfnamefont {C.}~\bibnamefont {Girit}}, \bibinfo {author} {\bibfnamefont {Z.}~\bibnamefont {Hao}}, \bibinfo {author} {\bibfnamefont {M.~C.}\ \bibnamefont {Martin}}, \bibinfo {author} {\bibfnamefont {A.}~\bibnamefont {Zettl}}, \bibinfo {author} {\bibfnamefont {M.~F.}\ \bibnamefont {Crommie}}, \bibinfo {author} {\bibfnamefont {Y.~R.}\ \bibnamefont {Shen}},\ and\ \bibinfo {author} {\bibfnamefont {F.}~\bibnamefont {Wang}},\ }\bibfield  {title} {\bibinfo {title} {Direct observation of a widely tunable bandgap in bilayer graphene},\ }\href {https://doi.org/10.1038/nature08105} {\bibfield  {journal} {\bibinfo  {journal} {Nature}\ }\textbf {\bibinfo {volume} {459}},\ \bibinfo {pages} {820–823} (\bibinfo {year} {2009})}\BibitemShut {NoStop}%
\bibitem [{\citenamefont {Kuzmenko}\ \emph {et~al.}(2009)\citenamefont {Kuzmenko}, \citenamefont {van Heumen}, \citenamefont {van~der Marel}, \citenamefont {Lerch}, \citenamefont {Blake}, \citenamefont {Novoselov},\ and\ \citenamefont {Geim}}]{Infrared_spectroscopy_of_electronic_bands_in_bilayer_graphene}%
  \BibitemOpen
  \bibfield  {author} {\bibinfo {author} {\bibfnamefont {A.~B.}\ \bibnamefont {Kuzmenko}}, \bibinfo {author} {\bibfnamefont {E.}~\bibnamefont {van Heumen}}, \bibinfo {author} {\bibfnamefont {D.}~\bibnamefont {van~der Marel}}, \bibinfo {author} {\bibfnamefont {P.}~\bibnamefont {Lerch}}, \bibinfo {author} {\bibfnamefont {P.}~\bibnamefont {Blake}}, \bibinfo {author} {\bibfnamefont {K.~S.}\ \bibnamefont {Novoselov}},\ and\ \bibinfo {author} {\bibfnamefont {A.~K.}\ \bibnamefont {Geim}},\ }\bibfield  {title} {\bibinfo {title} {Infrared spectroscopy of electronic bands in bilayer graphene},\ }\href {https://doi.org/10.1103/PhysRevB.79.115441} {\bibfield  {journal} {\bibinfo  {journal} {Phys. Rev. B}\ }\textbf {\bibinfo {volume} {79}},\ \bibinfo {pages} {115441} (\bibinfo {year} {2009})}\BibitemShut {NoStop}%
\bibitem [{\citenamefont {Wang}\ \emph {et~al.}(2023)\citenamefont {Wang}, \citenamefont {Zheng}, \citenamefont {Sun}, \citenamefont {Zhao}, \citenamefont {Zhang}, \citenamefont {Liu}, \citenamefont {Yang}, \citenamefont {Zhu}, \citenamefont {Shao}, \citenamefont {Wei}, \citenamefont {Yin}, \citenamefont {Yin}, \citenamefont {Lin}, \citenamefont {Jia},\ and\ \citenamefont {Liu}}]{Controlled_Growth_of_Bilayer_Graphene_on_Space_Confined_Cu_Substrates}%
  \BibitemOpen
  \bibfield  {author} {\bibinfo {author} {\bibfnamefont {S.}~\bibnamefont {Wang}}, \bibinfo {author} {\bibfnamefont {C.}~\bibnamefont {Zheng}}, \bibinfo {author} {\bibfnamefont {X.}~\bibnamefont {Sun}}, \bibinfo {author} {\bibfnamefont {Y.}~\bibnamefont {Zhao}}, \bibinfo {author} {\bibfnamefont {S.}~\bibnamefont {Zhang}}, \bibinfo {author} {\bibfnamefont {Q.}~\bibnamefont {Liu}}, \bibinfo {author} {\bibfnamefont {J.}~\bibnamefont {Yang}}, \bibinfo {author} {\bibfnamefont {Y.}~\bibnamefont {Zhu}}, \bibinfo {author} {\bibfnamefont {J.}~\bibnamefont {Shao}}, \bibinfo {author} {\bibfnamefont {M.}~\bibnamefont {Wei}}, \bibinfo {author} {\bibfnamefont {W.}~\bibnamefont {Yin}}, \bibinfo {author} {\bibfnamefont {J.}~\bibnamefont {Yin}}, \bibinfo {author} {\bibfnamefont {L.}~\bibnamefont {Lin}}, \bibinfo {author} {\bibfnamefont {K.}~\bibnamefont {Jia}},\ and\ \bibinfo {author} {\bibfnamefont {Z.}~\bibnamefont {Liu}},\ }\bibfield  {title} {\bibinfo {title} {Controlled growth of bilayer graphene on space-confined cu
  substrates},\ }\href {https://doi.org/10.1021/acs.jpcc.3c01586} {\bibfield  {journal} {\bibinfo  {journal} {The Journal of Physical Chemistry C}\ }\textbf {\bibinfo {volume} {127}},\ \bibinfo {pages} {13601} (\bibinfo {year} {2023})},\ \Eprint {https://arxiv.org/abs/https://doi.org/10.1021/acs.jpcc.3c01586} {https://doi.org/10.1021/acs.jpcc.3c01586} \BibitemShut {NoStop}%
\bibitem [{\citenamefont {Zhang}\ \emph {et~al.}(2024{\natexlab{a}})\citenamefont {Zhang}, \citenamefont {Zhou}, \citenamefont {Ren}, \citenamefont {Feng}, \citenamefont {Qiao}, \citenamefont {Wang}, \citenamefont {Wang}, \citenamefont {Bai}, \citenamefont {Wu}, \citenamefont {Tang}, \citenamefont {Zhou}, \citenamefont {Liu},\ and\ \citenamefont {Xu}}]{Towards_growth_of_pure_AB_stacked_bilayer_graphene_single_crystals}%
  \BibitemOpen
  \bibfield  {author} {\bibinfo {author} {\bibfnamefont {X.}~\bibnamefont {Zhang}}, \bibinfo {author} {\bibfnamefont {T.}~\bibnamefont {Zhou}}, \bibinfo {author} {\bibfnamefont {Y.}~\bibnamefont {Ren}}, \bibinfo {author} {\bibfnamefont {Z.}~\bibnamefont {Feng}}, \bibinfo {author} {\bibfnamefont {R.}~\bibnamefont {Qiao}}, \bibinfo {author} {\bibfnamefont {Q.}~\bibnamefont {Wang}}, \bibinfo {author} {\bibfnamefont {B.}~\bibnamefont {Wang}}, \bibinfo {author} {\bibfnamefont {J.}~\bibnamefont {Bai}}, \bibinfo {author} {\bibfnamefont {M.}~\bibnamefont {Wu}}, \bibinfo {author} {\bibfnamefont {Z.}~\bibnamefont {Tang}}, \bibinfo {author} {\bibfnamefont {X.}~\bibnamefont {Zhou}}, \bibinfo {author} {\bibfnamefont {K.}~\bibnamefont {Liu}},\ and\ \bibinfo {author} {\bibfnamefont {X.}~\bibnamefont {Xu}},\ }\bibfield  {title} {\bibinfo {title} {Towards growth of pure ab-stacked bilayer graphene single crystals},\ }\href {https://doi.org/https://doi.org/10.1007/s12274-023-6348-9} {\bibfield  {journal} {\bibinfo  {journal}
  {Nano Research}\ }\textbf {\bibinfo {volume} {17}},\ \bibinfo {pages} {4616–4621} (\bibinfo {year} {2024}{\natexlab{a}})}\BibitemShut {NoStop}%
\bibitem [{\citenamefont {Zhang}\ \emph {et~al.}(2024{\natexlab{b}})\citenamefont {Zhang}, \citenamefont {Zhou}, \citenamefont {Chen}, \citenamefont {Jaroš}, \citenamefont {Kolíbal}, \citenamefont {Bábor}, \citenamefont {Zhang}, \citenamefont {Yan}, \citenamefont {Qiao}, \citenamefont {Zhang}, \citenamefont {Zhang}, \citenamefont {Wei}, \citenamefont {Cui}, \citenamefont {Qiao}, \citenamefont {Liu}, \citenamefont {Bao}, \citenamefont {Yang}, \citenamefont {Cheng}, \citenamefont {Wang}, \citenamefont {Wang}, \citenamefont {Liu}, \citenamefont {Willinger}, \citenamefont {Gao}, \citenamefont {Liu}, \citenamefont {Ji},\ and\ \citenamefont {Wang}}]{Layer_by_layer_growth_of_bilayer_graphene_single_crystals_enabled_by_proximity_catalytic_activity}%
  \BibitemOpen
  \bibfield  {author} {\bibinfo {author} {\bibfnamefont {Z.}~\bibnamefont {Zhang}}, \bibinfo {author} {\bibfnamefont {L.}~\bibnamefont {Zhou}}, \bibinfo {author} {\bibfnamefont {Z.}~\bibnamefont {Chen}}, \bibinfo {author} {\bibfnamefont {A.}~\bibnamefont {Jaroš}}, \bibinfo {author} {\bibfnamefont {M.}~\bibnamefont {Kolíbal}}, \bibinfo {author} {\bibfnamefont {P.}~\bibnamefont {Bábor}}, \bibinfo {author} {\bibfnamefont {Q.}~\bibnamefont {Zhang}}, \bibinfo {author} {\bibfnamefont {C.}~\bibnamefont {Yan}}, \bibinfo {author} {\bibfnamefont {R.}~\bibnamefont {Qiao}}, \bibinfo {author} {\bibfnamefont {Q.}~\bibnamefont {Zhang}}, \bibinfo {author} {\bibfnamefont {T.}~\bibnamefont {Zhang}}, \bibinfo {author} {\bibfnamefont {W.}~\bibnamefont {Wei}}, \bibinfo {author} {\bibfnamefont {Y.}~\bibnamefont {Cui}}, \bibinfo {author} {\bibfnamefont {J.}~\bibnamefont {Qiao}}, \bibinfo {author} {\bibfnamefont {L.}~\bibnamefont {Liu}}, \bibinfo {author} {\bibfnamefont {L.}~\bibnamefont {Bao}}, \bibinfo {author} {\bibfnamefont
  {H.}~\bibnamefont {Yang}}, \bibinfo {author} {\bibfnamefont {Z.}~\bibnamefont {Cheng}}, \bibinfo {author} {\bibfnamefont {Y.}~\bibnamefont {Wang}}, \bibinfo {author} {\bibfnamefont {E.}~\bibnamefont {Wang}}, \bibinfo {author} {\bibfnamefont {Z.}~\bibnamefont {Liu}}, \bibinfo {author} {\bibfnamefont {M.}~\bibnamefont {Willinger}}, \bibinfo {author} {\bibfnamefont {H.-J.}\ \bibnamefont {Gao}}, \bibinfo {author} {\bibfnamefont {K.}~\bibnamefont {Liu}}, \bibinfo {author} {\bibfnamefont {W.}~\bibnamefont {Ji}},\ and\ \bibinfo {author} {\bibfnamefont {Z.-J.}\ \bibnamefont {Wang}},\ }\bibfield  {title} {\bibinfo {title} {Layer-by-layer growth of bilayer graphene single-crystals enabled by proximity catalytic activity},\ }\href {https://doi.org/https://doi.org/10.1016/j.nantod.2024.102482} {\bibfield  {journal} {\bibinfo  {journal} {Nano Today}\ }\textbf {\bibinfo {volume} {59}},\ \bibinfo {pages} {102482} (\bibinfo {year} {2024}{\natexlab{b}})}\BibitemShut {NoStop}%
\bibitem [{\citenamefont {Tang}\ \emph {et~al.}(2025)\citenamefont {Tang}, \citenamefont {Ma}, \citenamefont {Ji}, \citenamefont {Han}, \citenamefont {Xu},\ and\ \citenamefont {Ren}}]{Uniform_growth_of_AB_stacked_bilayer_graphene_single_crystal_by_cyclic_segregation/dissolution}%
  \BibitemOpen
  \bibfield  {author} {\bibinfo {author} {\bibfnamefont {Y.}~\bibnamefont {Tang}}, \bibinfo {author} {\bibfnamefont {L.-P.}\ \bibnamefont {Ma}}, \bibinfo {author} {\bibfnamefont {K.}~\bibnamefont {Ji}}, \bibinfo {author} {\bibfnamefont {T.}~\bibnamefont {Han}}, \bibinfo {author} {\bibfnamefont {J.}~\bibnamefont {Xu}},\ and\ \bibinfo {author} {\bibfnamefont {W.}~\bibnamefont {Ren}},\ }\bibfield  {title} {\bibinfo {title} {Uniform growth of ab-stacked bilayer graphene single crystal by cyclic segregation/dissolution},\ }\href {https://doi.org/https://doi.org/10.1016/j.carbon.2025.120317} {\bibfield  {journal} {\bibinfo  {journal} {Carbon}\ }\textbf {\bibinfo {volume} {239}},\ \bibinfo {pages} {120317} (\bibinfo {year} {2025})}\BibitemShut {NoStop}%
\bibitem [{\citenamefont {Yin}\ \emph {et~al.}(2022)\citenamefont {Yin}, \citenamefont {Tan}, \citenamefont {Barcons-Ruiz}, \citenamefont {Torre}, \citenamefont {Watanabe}, \citenamefont {Taniguchi}, \citenamefont {Song}, \citenamefont {Hone},\ and\ \citenamefont {Koppens}}]{Tunable_and_giant_valley_selective_Hall_effect_in_gapped_bilayer_graphene_yin_tan_barcons-ruiz_torre_watanabe_taniguchi_song_hone_koppens_2022}%
  \BibitemOpen
  \bibfield  {author} {\bibinfo {author} {\bibfnamefont {J.}~\bibnamefont {Yin}}, \bibinfo {author} {\bibfnamefont {C.}~\bibnamefont {Tan}}, \bibinfo {author} {\bibfnamefont {D.}~\bibnamefont {Barcons-Ruiz}}, \bibinfo {author} {\bibfnamefont {I.}~\bibnamefont {Torre}}, \bibinfo {author} {\bibfnamefont {K.}~\bibnamefont {Watanabe}}, \bibinfo {author} {\bibfnamefont {T.}~\bibnamefont {Taniguchi}}, \bibinfo {author} {\bibfnamefont {J.~C.~W.}\ \bibnamefont {Song}}, \bibinfo {author} {\bibfnamefont {J.}~\bibnamefont {Hone}},\ and\ \bibinfo {author} {\bibfnamefont {F.~H.~L.}\ \bibnamefont {Koppens}},\ }\bibfield  {title} {\bibinfo {title} {Tunable and giant valley-selective {Hall} effect in gapped bilayer graphene},\ }\href {https://doi.org/10.1126/science.abl4266} {\bibfield  {journal} {\bibinfo  {journal} {Science}\ }\textbf {\bibinfo {volume} {375}},\ \bibinfo {pages} {1398–1402} (\bibinfo {year} {2022})}\BibitemShut {NoStop}%
\bibitem [{\citenamefont {Oka}\ and\ \citenamefont {Bucciantini}(2016)}]{Heterodyne_Hall_effect_in_a_two_dimensional_electron_gas}%
  \BibitemOpen
  \bibfield  {author} {\bibinfo {author} {\bibfnamefont {T.}~\bibnamefont {Oka}}\ and\ \bibinfo {author} {\bibfnamefont {L.}~\bibnamefont {Bucciantini}},\ }\bibfield  {title} {\bibinfo {title} {Heterodyne hall effect in a two-dimensional electron gas},\ }\href {https://doi.org/10.1103/PhysRevB.94.155133} {\bibfield  {journal} {\bibinfo  {journal} {Phys. Rev. B}\ }\textbf {\bibinfo {volume} {94}},\ \bibinfo {pages} {155133} (\bibinfo {year} {2016})}\BibitemShut {NoStop}%
\bibitem [{\citenamefont {Peralta~Gavensky}\ \emph {et~al.}(2018)\citenamefont {Peralta~Gavensky}, \citenamefont {Usaj},\ and\ \citenamefont {Balseiro}}]{Time-resolved_Hall_conductivity_of_pulse_driven_topological_quantum_systems}%
  \BibitemOpen
  \bibfield  {author} {\bibinfo {author} {\bibfnamefont {L.}~\bibnamefont {Peralta~Gavensky}}, \bibinfo {author} {\bibfnamefont {G.}~\bibnamefont {Usaj}},\ and\ \bibinfo {author} {\bibfnamefont {C.~A.}\ \bibnamefont {Balseiro}},\ }\bibfield  {title} {\bibinfo {title} {Time-resolved hall conductivity of pulse-driven topological quantum systems},\ }\href {https://doi.org/10.1103/PhysRevB.98.165414} {\bibfield  {journal} {\bibinfo  {journal} {Phys. Rev. B}\ }\textbf {\bibinfo {volume} {98}},\ \bibinfo {pages} {165414} (\bibinfo {year} {2018})}\BibitemShut {NoStop}%
\bibitem [{sup()}]{suppmaterial}%
  \BibitemOpen
  \href@noop {} {}\bibinfo {note} {See Supplemental Material at [URL will be inserted by publisher] for a detailed description of the kernel polynomial method, a derivation of the formalism used to obtain the out-of-equilibrium Kubo formula for the electrical conductivity, the time profile of the light-induced Hall conductivity for different thermalization times, and the formula to compute the fluence of the incident light pulse.}\BibitemShut {Stop}%
\bibitem [{\citenamefont {Massicotte}\ \emph {et~al.}(2021)\citenamefont {Massicotte}, \citenamefont {Soavi}, \citenamefont {Principi},\ and\ \citenamefont {Tielrooij}}]{Hot_carriers_in_graphene_fundamentals_and_applications_massicotte_giancarlo_soavi_principi_klaas-jan_tielrooij_2021}%
  \BibitemOpen
  \bibfield  {author} {\bibinfo {author} {\bibfnamefont {M.}~\bibnamefont {Massicotte}}, \bibinfo {author} {\bibfnamefont {G.}~\bibnamefont {Soavi}}, \bibinfo {author} {\bibfnamefont {A.}~\bibnamefont {Principi}},\ and\ \bibinfo {author} {\bibfnamefont {K.-J.}\ \bibnamefont {Tielrooij}},\ }\bibfield  {title} {\bibinfo {title} {Hot carriers in graphene – fundamentals and applications},\ }\href {https://doi.org/10.1039/d0nr09166a} {\bibfield  {journal} {\bibinfo  {journal} {Nanoscale}\ }\textbf {\bibinfo {volume} {13}},\ \bibinfo {pages} {8376–8411} (\bibinfo {year} {2021})}\BibitemShut {NoStop}%
\bibitem [{\citenamefont {Aversa}\ and\ \citenamefont {Sipe}(1995)}]{Nonlinear_optical_susceptibilities_of_semiconductors_Results_with_a_length_gauge_analysis_aversa_sipe_1995}%
  \BibitemOpen
  \bibfield  {author} {\bibinfo {author} {\bibfnamefont {C.}~\bibnamefont {Aversa}}\ and\ \bibinfo {author} {\bibfnamefont {J.~E.}\ \bibnamefont {Sipe}},\ }\bibfield  {title} {\bibinfo {title} {Nonlinear optical susceptibilities of semiconductors: Results with a length-gauge analysis},\ }\href {https://doi.org/10.1103/physrevb.52.14636} {\bibfield  {journal} {\bibinfo  {journal} {Physical Review B}\ }\textbf {\bibinfo {volume} {52}},\ \bibinfo {pages} {14636–14645} (\bibinfo {year} {1995})}\BibitemShut {NoStop}%
\bibitem [{\citenamefont {Di~Ventra}(2008)}]{Electrical_Transport_in_Nanoscale_Systems_Di_Ventra_2008}%
  \BibitemOpen
  \bibfield  {author} {\bibinfo {author} {\bibfnamefont {M.}~\bibnamefont {Di~Ventra}},\ }\href@noop {} {\emph {\bibinfo {title} {Electrical Transport in Nanoscale Systems}}}\ (\bibinfo  {publisher} {Cambridge University Press},\ \bibinfo {year} {2008})\BibitemShut {NoStop}%
\bibitem [{\citenamefont {Weiße}\ \emph {et~al.}(2006)\citenamefont {Weiße}, \citenamefont {Wellein}, \citenamefont {Alvermann},\ and\ \citenamefont {Fehske}}]{The_kernel_polynomial_method_weiße_wellein_alvermann_fehske_2006}%
  \BibitemOpen
  \bibfield  {author} {\bibinfo {author} {\bibfnamefont {A.}~\bibnamefont {Weiße}}, \bibinfo {author} {\bibfnamefont {G.}~\bibnamefont {Wellein}}, \bibinfo {author} {\bibfnamefont {A.}~\bibnamefont {Alvermann}},\ and\ \bibinfo {author} {\bibfnamefont {H.}~\bibnamefont {Fehske}},\ }\bibfield  {title} {\bibinfo {title} {The kernel polynomial method},\ }\href {https://doi.org/10.1103/revmodphys.78.275} {\bibfield  {journal} {\bibinfo  {journal} {Reviews of Modern Physics}\ }\textbf {\bibinfo {volume} {78}},\ \bibinfo {pages} {275–306} (\bibinfo {year} {2006})}\BibitemShut {NoStop}%
\bibitem [{\citenamefont {Calvo}\ \emph {et~al.}(2011)\citenamefont {Calvo}, \citenamefont {Pastawski}, \citenamefont {Roche},\ and\ \citenamefont {Foa~Torres}}]{Tuning_laser_induced_band_gaps_in_graphene_calvo_pastawski_roche_luis_2011}%
  \BibitemOpen
  \bibfield  {author} {\bibinfo {author} {\bibfnamefont {H.~L.}\ \bibnamefont {Calvo}}, \bibinfo {author} {\bibfnamefont {H.~M.}\ \bibnamefont {Pastawski}}, \bibinfo {author} {\bibfnamefont {S.}~\bibnamefont {Roche}},\ and\ \bibinfo {author} {\bibfnamefont {L.~E.~F.}\ \bibnamefont {Foa~Torres}},\ }\bibfield  {title} {\bibinfo {title} {Tuning laser-induced band gaps in graphene},\ }\href {https://doi.org/10.1063/1.3597412} {\bibfield  {journal} {\bibinfo  {journal} {Applied Physics Letters}\ }\textbf {\bibinfo {volume} {98}},\ \bibinfo {pages} {232103} (\bibinfo {year} {2011})}\BibitemShut {NoStop}%
\bibitem [{\citenamefont {Keller}(2021)}]{Ultrafast_Lasers_keller_2021}%
  \BibitemOpen
  \bibfield  {author} {\bibinfo {author} {\bibfnamefont {U.}~\bibnamefont {Keller}},\ }\href {https://doi.org/10.1007/978-3-030-82532-4} {\emph {\bibinfo {title} {Ultrafast Lasers}}}\ (\bibinfo  {publisher} {Springer International Publishing},\ \bibinfo {year} {2021})\BibitemShut {NoStop}%
\bibitem [{\citenamefont {Ghalamkari}\ \emph {et~al.}(2018)\citenamefont {Ghalamkari}, \citenamefont {Tatsumi},\ and\ \citenamefont {Saito}}]{Perfect_Circular_Dichroism_in_the_Haldane_Model_kazu_ghalamkari_tatsumi_saito_2018}%
  \BibitemOpen
  \bibfield  {author} {\bibinfo {author} {\bibfnamefont {K.}~\bibnamefont {Ghalamkari}}, \bibinfo {author} {\bibfnamefont {Y.}~\bibnamefont {Tatsumi}},\ and\ \bibinfo {author} {\bibfnamefont {R.}~\bibnamefont {Saito}},\ }\bibfield  {title} {\bibinfo {title} {Perfect circular dichroism in the haldane model},\ }\href {https://doi.org/10.7566/jpsj.87.063708} {\bibfield  {journal} {\bibinfo  {journal} {Journal of the Physical Society of Japan}\ }\textbf {\bibinfo {volume} {87}},\ \bibinfo {pages} {063708} (\bibinfo {year} {2018})}\BibitemShut {NoStop}%
\bibitem [{\citenamefont {McCann}\ and\ \citenamefont {Koshino}(2013)}]{The_electronic_properties_of_bilayer_graphene_mccann_koshino_2013}%
  \BibitemOpen
  \bibfield  {author} {\bibinfo {author} {\bibfnamefont {E.}~\bibnamefont {McCann}}\ and\ \bibinfo {author} {\bibfnamefont {M.}~\bibnamefont {Koshino}},\ }\bibfield  {title} {\bibinfo {title} {The electronic properties of bilayer graphene},\ }\href {https://doi.org/10.1088/0034-4885/76/5/056503} {\bibfield  {journal} {\bibinfo  {journal} {Reports on Progress in Physics}\ }\textbf {\bibinfo {volume} {76}},\ \bibinfo {pages} {056503} (\bibinfo {year} {2013})}\BibitemShut {NoStop}%
\bibitem [{\citenamefont {Tuan}\ \emph {et~al.}(2016)\citenamefont {Tuan}, \citenamefont {Adam},\ and\ \citenamefont {Roche}}]{Spin_dynamics_in_bilayer_graphene_Role_of_electron-hole_puddles_and_Dyakonov-Perel_mechanism_tuan_adam_roche_2016}%
  \BibitemOpen
  \bibfield  {author} {\bibinfo {author} {\bibfnamefont {D.~V.}\ \bibnamefont {Tuan}}, \bibinfo {author} {\bibfnamefont {S.}~\bibnamefont {Adam}},\ and\ \bibinfo {author} {\bibfnamefont {S.}~\bibnamefont {Roche}},\ }\bibfield  {title} {\bibinfo {title} {Spin dynamics in bilayer graphene: Role of electron-hole puddles and dyakonov-perel mechanism},\ }\href {https://doi.org/10.1103/physrevb.94.041405} {\bibfield  {journal} {\bibinfo  {journal} {Physical Review B}\ }\textbf {\bibinfo {volume} {94}},\ \bibinfo {pages} {041405} (\bibinfo {year} {2016})}\BibitemShut {NoStop}%
\bibitem [{\citenamefont {Cupo}\ \emph {et~al.}(2025)\citenamefont {Cupo}, \citenamefont {Cheng}, \citenamefont {Ramanathan},\ and\ \citenamefont {Viola}}]{Generation_of_Ultrahigh_Anomalous_Hall_Conductivities_via_Optimally_Prepared_Topological_Floquet_States}%
  \BibitemOpen
  \bibfield  {author} {\bibinfo {author} {\bibfnamefont {A.}~\bibnamefont {Cupo}}, \bibinfo {author} {\bibfnamefont {H.-P.}\ \bibnamefont {Cheng}}, \bibinfo {author} {\bibfnamefont {C.}~\bibnamefont {Ramanathan}},\ and\ \bibinfo {author} {\bibfnamefont {L.}~\bibnamefont {Viola}},\ }\href {https://arxiv.org/abs/2511.19843} {\bibinfo {title} {Generation of ultrahigh anomalous hall conductivities via optimally prepared topological floquet states}} (\bibinfo {year} {2025}),\ \Eprint {https://arxiv.org/abs/2511.19843} {arXiv:2511.19843 [cond-mat.mes-hall]} \BibitemShut {NoStop}%
\bibitem [{\citenamefont {Martin}\ \emph {et~al.}(2007)\citenamefont {Martin}, \citenamefont {Akerman}, \citenamefont {Ulbricht}, \citenamefont {Lohmann}, \citenamefont {Smet}, \citenamefont {von Klitzing},\ and\ \citenamefont {Yacoby}}]{Observation_of_electron_hole_puddles_in_graphene_using_a_scanning_single_electron_transistor_martin_akerman_ulbricht_lohmann_smet_von_klitzing_yacoby_2007}%
  \BibitemOpen
  \bibfield  {author} {\bibinfo {author} {\bibfnamefont {J.}~\bibnamefont {Martin}}, \bibinfo {author} {\bibfnamefont {N.}~\bibnamefont {Akerman}}, \bibinfo {author} {\bibfnamefont {G.}~\bibnamefont {Ulbricht}}, \bibinfo {author} {\bibfnamefont {T.}~\bibnamefont {Lohmann}}, \bibinfo {author} {\bibfnamefont {J.~H.}\ \bibnamefont {Smet}}, \bibinfo {author} {\bibfnamefont {K.}~\bibnamefont {von Klitzing}},\ and\ \bibinfo {author} {\bibfnamefont {A.}~\bibnamefont {Yacoby}},\ }\bibfield  {title} {\bibinfo {title} {Observation of electron–hole puddles in graphene using a scanning single-electron transistor},\ }\href {https://doi.org/10.1038/nphys781} {\bibfield  {journal} {\bibinfo  {journal} {Nature Physics}\ }\textbf {\bibinfo {volume} {4}},\ \bibinfo {pages} {144–148} (\bibinfo {year} {2007})}\BibitemShut {NoStop}%
\bibitem [{\citenamefont {Deshpande}\ \emph {et~al.}(2009)\citenamefont {Deshpande}, \citenamefont {Bao}, \citenamefont {Miao}, \citenamefont {Lau},\ and\ \citenamefont {LeRoy}}]{Spatially_resolved_spectroscopy_of_monolayer_graphene_onSiO2_deshpande_bao_miao_lau_leroy_2009}%
  \BibitemOpen
  \bibfield  {author} {\bibinfo {author} {\bibfnamefont {A.}~\bibnamefont {Deshpande}}, \bibinfo {author} {\bibfnamefont {W.}~\bibnamefont {Bao}}, \bibinfo {author} {\bibfnamefont {F.}~\bibnamefont {Miao}}, \bibinfo {author} {\bibfnamefont {C.~N.}\ \bibnamefont {Lau}},\ and\ \bibinfo {author} {\bibfnamefont {B.~J.}\ \bibnamefont {LeRoy}},\ }\bibfield  {title} {\bibinfo {title} {{Spatially resolved spectroscopy of monolayer graphene on SiO$_2$}},\ }\href {https://doi.org/10.1103/PhysRevB.79.205411} {\bibfield  {journal} {\bibinfo  {journal} {Physical Review B}\ }\textbf {\bibinfo {volume} {79}},\ \bibinfo {pages} {205411} (\bibinfo {year} {2009})}\BibitemShut {NoStop}%
\bibitem [{\citenamefont {Adam}\ \emph {et~al.}(2011)\citenamefont {Adam}, \citenamefont {Jung}, \citenamefont {Klimov}, \citenamefont {Zhitenev}, \citenamefont {Stroscio},\ and\ \citenamefont {Stiles}}]{Mechanism_for_puddle_formation_in_graphene_adam_jung_klimov_zhitenev_stroscio_stiles_2011}%
  \BibitemOpen
  \bibfield  {author} {\bibinfo {author} {\bibfnamefont {S.}~\bibnamefont {Adam}}, \bibinfo {author} {\bibfnamefont {S.}~\bibnamefont {Jung}}, \bibinfo {author} {\bibfnamefont {N.~N.}\ \bibnamefont {Klimov}}, \bibinfo {author} {\bibfnamefont {N.~B.}\ \bibnamefont {Zhitenev}}, \bibinfo {author} {\bibfnamefont {J.~A.}\ \bibnamefont {Stroscio}},\ and\ \bibinfo {author} {\bibfnamefont {M.~D.}\ \bibnamefont {Stiles}},\ }\bibfield  {title} {\bibinfo {title} {Mechanism for puddle formation in graphene},\ }\href {https://doi.org/10.1103/physrevb.84.235421} {\bibfield  {journal} {\bibinfo  {journal} {Physical Review B}\ }\textbf {\bibinfo {volume} {84}},\ \bibinfo {pages} {235421} (\bibinfo {year} {2011})}\BibitemShut {NoStop}%
\bibitem [{\citenamefont {Rutter}\ \emph {et~al.}(2011)\citenamefont {Rutter}, \citenamefont {Jung}, \citenamefont {Klimov}, \citenamefont {Newell}, \citenamefont {Zhitenev},\ and\ \citenamefont {Stroscio}}]{Microscopic_polarization_in_bilayer_graphene_rutter_jung_klimov_newell_zhitenev_stroscio_2011}%
  \BibitemOpen
  \bibfield  {author} {\bibinfo {author} {\bibfnamefont {G.~M.}\ \bibnamefont {Rutter}}, \bibinfo {author} {\bibfnamefont {S.}~\bibnamefont {Jung}}, \bibinfo {author} {\bibfnamefont {N.~N.}\ \bibnamefont {Klimov}}, \bibinfo {author} {\bibfnamefont {D.~B.}\ \bibnamefont {Newell}}, \bibinfo {author} {\bibfnamefont {N.~B.}\ \bibnamefont {Zhitenev}},\ and\ \bibinfo {author} {\bibfnamefont {J.~A.}\ \bibnamefont {Stroscio}},\ }\bibfield  {title} {\bibinfo {title} {Microscopic polarization in bilayer graphene},\ }\href {https://doi.org/10.1038/nphys1988} {\bibfield  {journal} {\bibinfo  {journal} {Nature Physics}\ }\textbf {\bibinfo {volume} {7}},\ \bibinfo {pages} {649–655} (\bibinfo {year} {2011})}\BibitemShut {NoStop}%
\end{thebibliography}
\end{document}